%% file: XtoZV2017.tex
\pdfoutput=1 %% for JHEP
\documentclass[cernpreprint,texlive=2016,txfonts,UKenglish]{latex/atlasdoc}

\usepackage[subfigure=true]{latex/atlaspackage}
\usepackage{latex/atlasbiblatex}
\usepackage{pifont}
\usepackage{xcolor,colortbl}

\usepackage{latex/atlasphysics}

\graphicspath{{logos/}{figures/}}

\usepackage{multirow}
\usepackage{XtoZV2017-defs}

\hypersetup{pdftitle={ATLAS draft},pdfauthor={The ATLAS Collaboration}}
\input{XtoZV2017-metadata}
\begin{document}

\maketitle

\clearpage
\section{Introduction}
\label{sec:intro}
\input{tex/introduction.tex}

\section{ATLAS detector}
\label{sec:detector}
\input{tex/detector.tex}

\section{Data, signal models and simulation}
\label{sec:samples}
\input{tex/samples.tex}

\section{Event reconstruction}
\label{sec:objects}
\input{tex/objects}

\section{$X\to ZV\to\ell\ell\,qq$ search}
\label{sec:llqq}
\input{tex/llqq.tex}

\section{$X\to ZV\to\nu\nu\,qq$ search}
\label{sec:vvqq}
\input{tex/vvqq.tex}

\section{Systematic uncertainties}
\label{sec:systematics}
\input{tex/systematics.tex}

\section{Results}
\label{sec:results}
\input{tex/results.tex}

\FloatBarrier
\section{Conclusion}
\label{sec:conclusion}
\input{tex/conclusion.tex}

\section*{Acknowledgements}

\input{acknowledgements/Acknowledgements}

\clearpage

\printbibliography

\clearpage
\newpage
\input{atlas_authlist}

\clearpage

\end{document}

%% file: XtoZV2017-metadata.tex
\AtlasTitle{Searches for heavy $ZZ$ and $ZW$ resonances in the $\ell\ell qq$ and $\nu\nu qq$ final states in $pp$ collisions at $\sqrt{s}=13$~TeV with the ATLAS detector}

\author{The ATLAS Collaboration}

\AtlasRefCode{EXOT-2016-29}

\AtlasNote{EXOT-2016-29}

\PreprintIdNumber{CERN-EP-2017-146}

\AtlasJournal{JHEP}
\AtlasJournalRef{JHEP 03 (2018) 009}
\AtlasDOI{DOI:10.1007/JHEP03(2018)009}

\AtlasAbstract{%
This paper reports searches for heavy resonances decaying into $ZZ$ or $ZW$ using data from proton--proton collisions at a centre-of-mass energy of $\sqrt{s}=13$~TeV. The data, corresponding to an integrated luminosity of \lumi, were recorded with the ATLAS detector in 2015 and 2016 at the Large Hadron Collider. The searches are performed in final states in which one $Z$ boson decays into either a pair of light charged leptons (electrons and muons) or a pair of neutrinos, and the associated $W$ boson or the other $Z$ boson decays hadronically. No evidence of the production of heavy resonances is observed. Upper bounds on the production cross sections of heavy resonances times their decay branching ratios to $ZZ$ or $ZW$ are derived in the mass range 300--5000~GeV within the context of Standard Model extensions with additional Higgs bosons, a heavy vector triplet or warped extra dimensions. Production through gluon--gluon fusion, Drell--Yan or vector-boson fusion are considered, depending on the assumed model.
}

%% file: tex/introduction.tex
The discovery of a Higgs boson $h$ with a mass of approximately 125~GeV in 2012~\cite{atlas:2012obs,cms:2012obs} 
represents a major milestone in the understanding of electroweak symmetry breaking. Subsequent studies~\cite{Aad:2015gba,Aad:2015mxa,Khachatryan:2014jba,Khachatryan:2014kca} have shown that the properties of the new particle are consistent with those of the Standard Model~(SM) Higgs boson. Nevertheless, the possibility that the particle is part of an extended Higgs sector or other extension of the SM cannot be ruled out. Many of these models, motivated by hierarchy and naturalness arguments~\cite{Haber:1984rc,ArkaniHamed:1998rs,Giudice:2008bi}, predict the existence of new heavy resonances decaying into dibosons. In models with an extended Higgs sector, such as the two-Higgs-doublet models (2HDM)~\cite{Branco:2011iw} and the electroweak-singlet model~\cite{Hill:1987ea}, a heavy spin-0 neutral Higgs boson~($H$) can decay into a pair of $Z$ bosons. In extended gauge models~\cite{Altarelli:1989ff}, a heavier version of the SM $W$ boson~($W'$) is predicted to decay into $ZW$, and in models with warped extra dimensions~\cite{Randall:1999ee,Davoudiasl:1999tf}, spin-2 Kaluza--Klein~(KK) excitations of the graviton~($\Graviton$) are expected to decay into $ZZ$.   

\begin{figure}[!h]
\centering
  \subfigure[gluon--gluon fusion]{\includegraphics[width=0.325\textwidth]{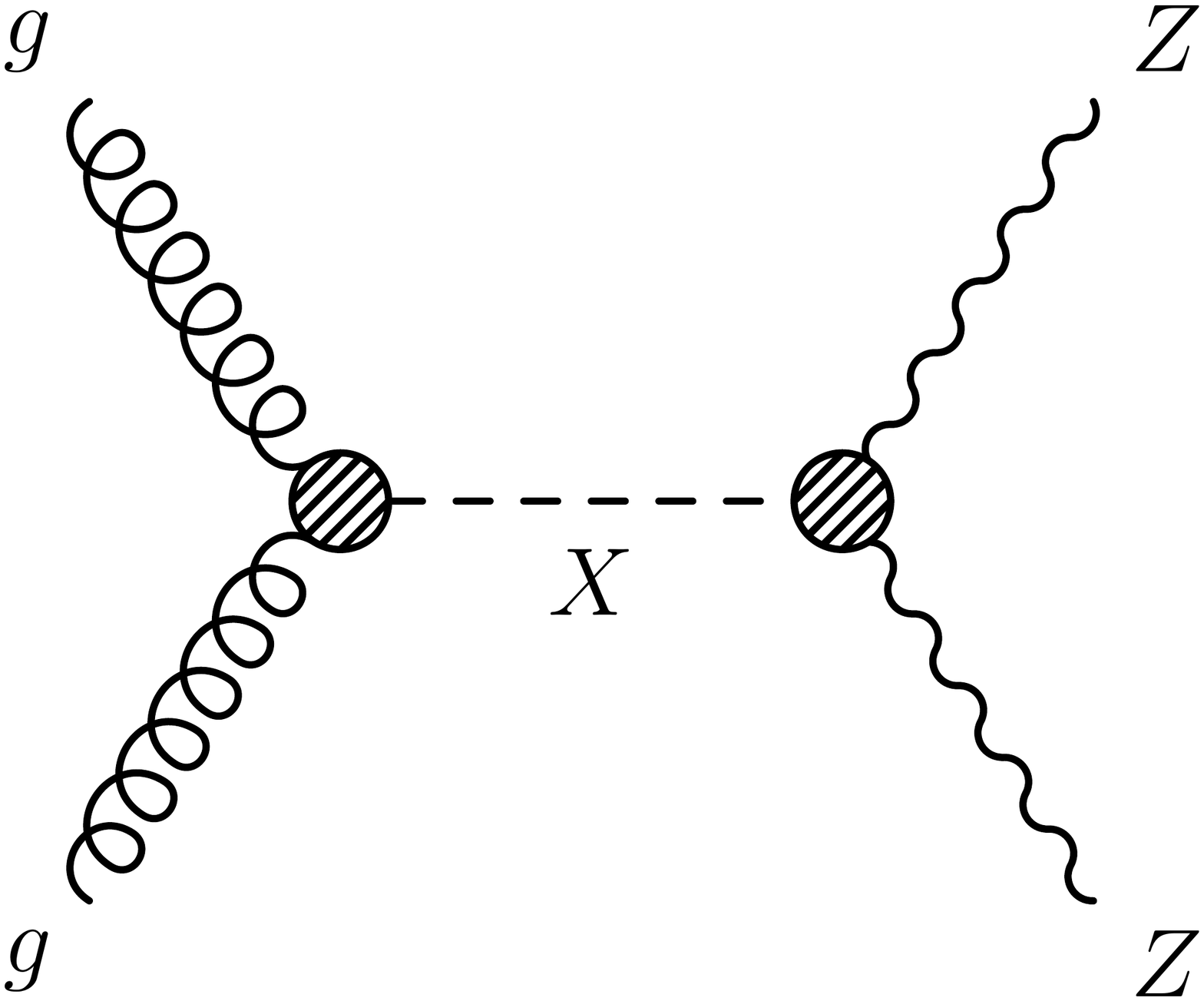}}
  \subfigure[Drell--Yan]{\includegraphics[width=0.325\textwidth]{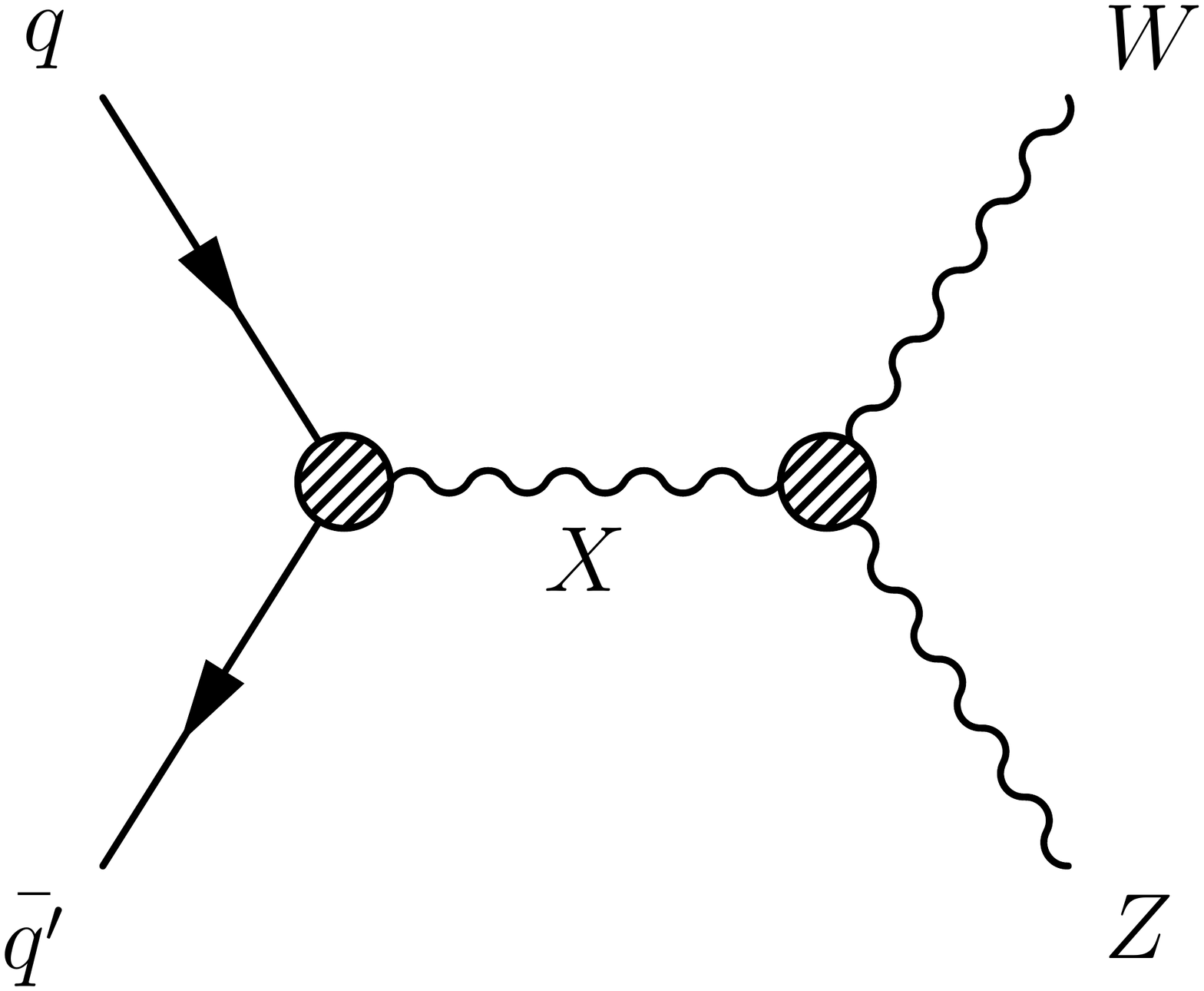}}
  \subfigure[vector-boson fusion]{\includegraphics[width=0.325\textwidth]{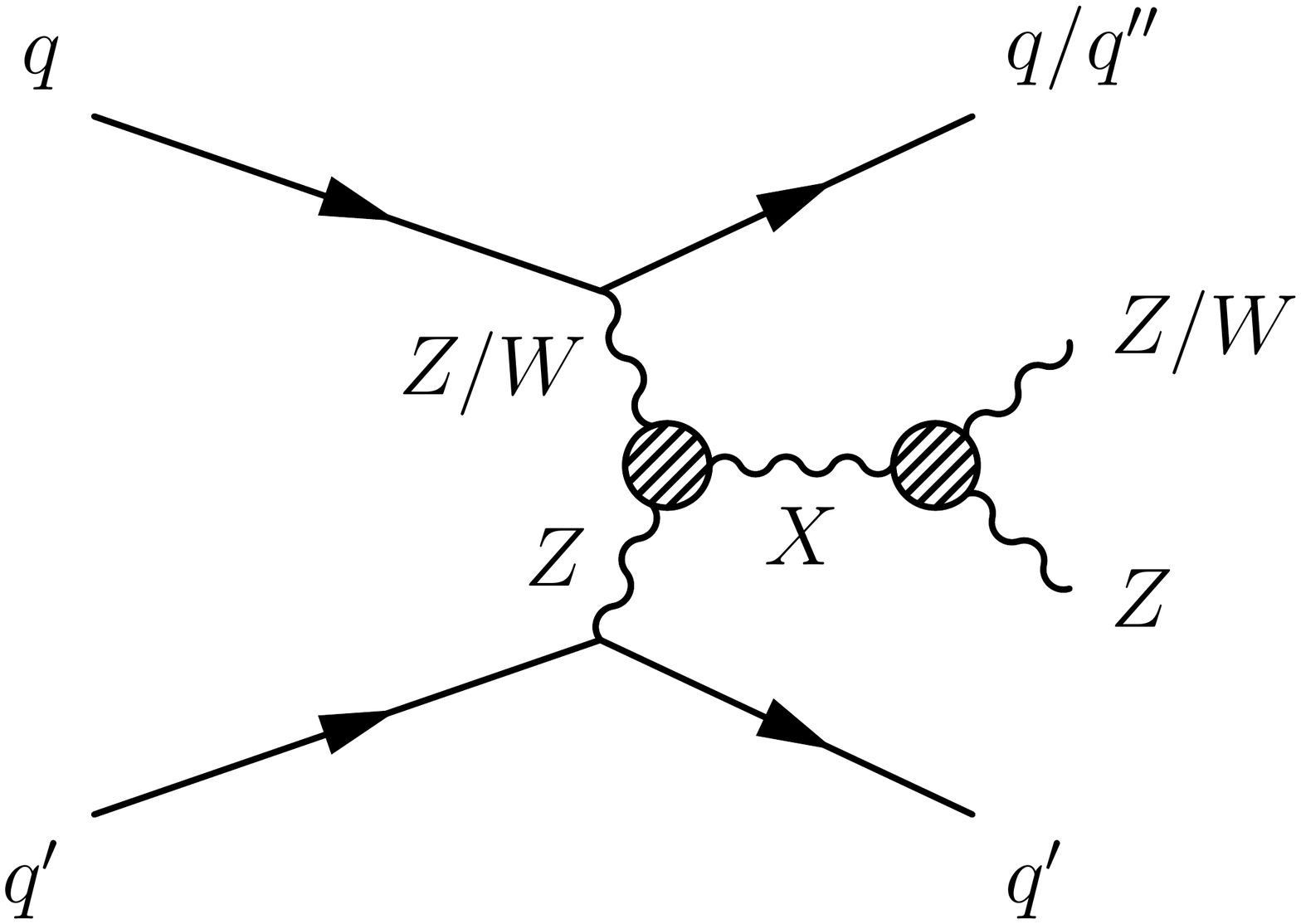}}
  \caption{\label{fig:diag} Representative Feynman diagrams for the production of heavy resonances $X$ with their decays into a pair of vector bosons.
}
\end{figure}

This paper reports searches for heavy resonances $X$ decaying into pairs of vector bosons, $ZV\,(V=W,Z)$. 
Production through gluon--gluon fusion~(ggF), Drell--Yan~(DY) and vector-boson fusion~(VBF) processes are considered, depending on the assumed model. Representative Feynman diagrams of these processes are shown in Figure~\ref{fig:diag}. Two $ZV$ decay modes are explored: one in which there is a $Z$ boson decaying into a pair of light charged leptons (electrons or muons, denoted by $\ell$), $Z\to\ell\ell$, \footnote{To simplify the notation, antiparticles are not explicitly labelled in this paper.} and the other in which a $Z$ boson decays into a pair of neutrinos, $Z\to\nu\nu$. In both cases, the vector boson $V$ is required to decay into a pair of quarks, $V\to qq$, leading to $X\to ZV\to\ell\ell qq$ and $X\to ZV\to\nu\nu qq$ decay modes.

Two different reconstruction techniques for the $V\to qq$ decay are considered: resolved and merged. The resolved reconstruction attempts to identify two separate small-radius jets (small-$R$ jet, or $j$) of hadrons from the $V\to qq$ decay, while the merged reconstruction uses jet substructure techniques to identify the $V\to qq$ decay reconstructed as a large-radius jet (large-$R$ jet or $J$). When the resonance mass is significantly higher than the $V$ boson mass, the $qq$ pair from the $V$ boson decay can be collimated. In this case, hadrons from the two quarks overlap in the detector and are more efficiently reconstructed as a single large-$R$ jet. The $X\to ZV\to\ell\ell qq$ searches utilise both reconstruction techniques for the $V\to qq$ decay, whereas the $X\to ZV\to\nu\nu qq$ search uses only the merged reconstruction.

Heavy resonances would manifest themselves as resonant structures above the SM background in the invariant-mass distributions of the $\ell\ell qq$ final state and as broad enhancements in the transverse-mass distributions of the $\nu\nu qq$ final state.  Thus for the $ZV\to\ell\ell qq$ decay mode, the invariant masses of the $\ell\ell J$ system ($m_{\ell\ell J}$) from the merged reconstruction and of the $\ell\ell jj$ system ($m_{\ell\ell jj}$) from the resolved reconstruction are used as the final discriminants for signal--background separation. For the $ZV\to\nu\nu qq$ decay mode, the final discriminant used is the transverse mass ($m_{\mathrm T}$) of the large-$R$ jet and the missing transverse momentum~(\met).

Both the ATLAS and CMS collaborations have reported searches for heavy resonances in $ZV$ decays in proton-proton ($pp$) collisions at $\sqrt{s}=8$~TeV and 13~TeV~\cite{Aad:2015kna,Aad:2015ipg,Sirunyan:2017nrt,Aaboud:2016okv,Aaboud:2017eta}. In addition to $\llqq$ and $\nu\nu qq$, these searches include fully hadronic ($qqqq$), semileptonic ($qq\ell\nu$), and fully leptonic ($\ell\ell\ell\ell,\, \ell\ell\ell\nu,\,\ell\ell\nu\nu$) final states. 
This paper extends previous ATLAS searches for $ZV$ resonances in the $\ell\ell qq$ and $\nu\nu qq$ final states at $\sqrt{s}=13$~TeV reported in Ref.~\cite{Aaboud:2016okv} and uses a dataset more than ten times larger.

%% file: tex/detector.tex
The ATLAS detector~\cite{atlasdet} at the Large Hadron Collider~ (LHC)~\cite{Evans:2008zzb} covers nearly the entire solid angle\footnote{ATLAS uses a right-handed coordinate system with its origin at the nominal interaction point (IP) in the centre of the detector and the $z$-axis along the beam pipe. The $x$-axis points from the IP to the centre of the LHC ring, and the $y$-axis points upward. Cylindrical coordinates $(r,\phi)$ are used in the transverse plane, $\phi$ being the azimuthal angle around the $z$-axis. The pseudorapidity is defined in terms of the polar angle $\theta$ as $\eta=-\ln\tan(\theta/2)$.} around the collision point,
and consists of an inner tracking detector surrounded by a thin superconducting solenoidal magnet producing
a 2\,T magnetic field, electromagnetic and hadronic calorimeters, and a muon spectrometer
incorporating three large toroid-magnet assemblies. The inner detector (ID) consists of a high-granularity silicon pixel
detector, including an insertable B-layer~\cite{ATLAS-TDR-19}, and a silicon microstrip tracker,
together providing precision tracking in the pseudorapidity
range $|\eta|<2.5$, complemented by a transition radiation tracker, providing tracking and electron identification
information for $|\eta|<2.0$. A lead/liquid-argon (LAr) electromagnetic calorimeter covers the region $|\eta|<3.2$,
and hadronic calorimetry is provided by steel/scintillator-tile calorimeters for $|\eta|<1.7$ and by copper/LAr
hadronic calorimeters for 1.7 < $|\eta|$ < 3.2. The forward region is covered by additional LAr calorimeters with copper and tungsten
absorbers. The muon spectrometer~(MS) consists of precision tracking chambers covering the region $|\eta|<2.7$,
and separate trigger chambers covering $|\eta|<2.4$. A two-level trigger system~\cite{Aaboud:2016leb} reduces the event rate to approximately 1~kHz for offline investigations.

%% file: tex/samples.tex
\subsection{Data}
\label{sec:data}
The data used in the searches were collected with the ATLAS detector in 2015 and 2016 $pp$ collisions at a centre-of-mass energy of 13~TeV, corresponding to a total integrated luminosity of~36.1~fb$^{-1}$.

\iffalse
Events used in the $X\to ZV\to\ell\ell qq$ search were recorded with a combination of multiple single-electron or single-muon triggers with varying transverse energy $\et$ (electron) and transverse momentum $\pt$ (muon), identification algorithm, and isolation requirements.  The identification and isolation criteria at trigger level are similar to those used in the offline reconstruction as described in Section~\ref{sec:objects}.
In the electron channel during the 2015 (2016) data-taking period, events were recorded by a trigger requiring at least one electron that satisfied $\et > 24$~\gev\ and the `medium'~(`tight') identification criteria, or $\et > 60$~\gev\ and the `medium' identification criteria, or $\et > 120~(140)$~\gev\ and the `loose' identification criteria. 
Events in the muon channel were recorded by a trigger requiring that at least one muon with $\pt > 20~(24)$~\gev\ with isolation requirement, or $\pt>50$~\gev\ without isolation requirement for the 2015~(2016) data-taking period. 
Events for the  $X\to ZV\to\nu\nu qq$ search were recorded using a \met\ trigger with a threshold of 70 GeV for the 2015 data and of 90~GeV or 110~GeV for the 2016 data.
\fi

Events used in the $X\to ZV\to\llqq$ search were recorded with a combination of multiple single-electron or single-muon triggers with varying transverse energy $\et$ (electron) and transverse momentum $\pt$ (muon), quality, and isolation requirements. The lowest $\et$ or $\pt$ requirement without trigger prescaling was 26~GeV for both the electrons and muons. Events for the $X\to ZV\to\nu\nu qq$ search were recorded with an \met\ trigger of varying threshold. The lowest threshold without prescaling is 100~GeV. This trigger is fully efficient for events passing the selection described below.

Events are retained for analysis if they were recorded with all detector systems operating normally and pass data-quality requirements~\cite{ATLAS-CONF-2015-029}.
Collision vertices are formed from tracks with $\pt>400\,\MeV$. 
If an event contains more than one vertex candidate, the one with the highest $\sum\pt^2$ of its associated tracks is selected as the primary vertex. All events are required to contain a primary vertex with at least two associated tracks.

\subsection{Signal models and simulation}
\label{sec:mc}
Three classes of models of physics beyond the Standard Model are used as benchmarks for the interpretation of the results. 
Different resonances are predicted by each of these:
a neutral heavy spin-0 Higgs boson $H$ using the narrow-width approximation~\cite{Barger:2007im}, a spin-1 $W'$ boson of the heavy vector triplet~(HVT) model~\cite{deBlas:2012qp,Pappadopulo:2014qza}, 
and a spin-2 KK graviton $\Graviton$ from the bulk Randall--Sundrum model~\cite{Randall:1999ee,Agashe:2007zd,Fitzpatrick}.
The new HVT bosons couple to the SM Higgs boson and gauge bosons with coupling strength $c_H g_V$ and to the SM fermions with coupling strength $(g^2/g_V )c_F$, where $g$ is the SM SU(2)$_{\mathrm L}$ coupling constant.
The parameter $g_V$ characterises the interaction strength of the new vector bosons, while the dimensionless coefficients $c_H$ and $c_F$ parameterise departures of this typical strength from interactions with the SM Higgs and gauge bosons and with fermions, respectively, and are expected to be of order unity in most models. In {\em Model A} with $g_V=1$, the branching
ratios of the new heavy vector boson to known fermions and gauge bosons are comparable, while in {\em Model B} with  $g_V=3$, fermionic couplings to the new heavy vector boson are suppressed, giving rise to larger branching ratios for decays into $ZW$ final states than in {\em Model A}. In a third model, {\em VBF Model}, 
the couplings to gauge bosons are similar to those in {\em Model A}, but the couplings to fermions are set to $0$.
In the bulk RS graviton model, the $\Graviton$ couplings to light fermions are suppressed and decays into final states involving heavy fermions, gauge bosons or Higgs bosons are favoured. The strength of the coupling depends on \kMPl, where $k$ corresponds to the curvature of the warped extra dimension and $\overline{M}_{\mathrm {Pl}}$ is the effective four-dimensional Planck scale $\overline{M}_{\mathrm{Pl}} = 2.4 \times 10^{18}$~\GeV{}. The cross section and intrinsic width scale as the square of \kMPl.

Monte Carlo (MC) samples of $H\to ZZ$ were generated by \textsc{Powheg-Box}~v1 \cite{Nason:2004rx, Frixione:2007vw, Alioli:2010xd,Alioli:2008tz} with the \textsc{CT10}~\cite{CT10} parton distribution functions~(PDF) assuming a Higgs boson with width far smaller than the experimental resolution. Both the ggF and VBF production processes are considered. 
Benchmark samples of HVT $W'\to ZW$ and  RS graviton $\Graviton\to ZZ$ were generated with \textsc{MadGraph5\_aMC@NLO}~2.2.2 ~\cite{Alwall:2014hca}, using the \textsc{NNPDF23LO} \cite{Ball:2012cx} PDF set. 
For the HVT model, two production modes of the charged vector triplet $W'$,  DY and VBF, are considered. 
The $W'$ resonance from DY production of {\em Model A} has a width approximately 2.6\% of its mass, while for VBF production the width is much narrower, since its couplings to fermions are set to $0$ in the VBF signal.
For HVT {\em Model B}, the resonance widths and experimental signatures are similar to those obtained for {\em Model A}. Thus results derived from {\em Model A} can be directly applied to benchmark {\em Model B} by rescaling the relevant branching ratios.
The $\Graviton$ has a mass-dependent width, 3.7\% at 500~GeV and 6.4\% at 5000~GeV relative to its mass, for $\kMPl=1$. An RS graviton with $\kMPl=0.5$ is also considered and the samples were obtained by reweighting the generated samples for $\kMPl=1$ to account for the resonance width and cross-section differences.
The parton showering and hadronisation were modelled with  \textsc{Pythia}~8.186 \cite{Sjostrand:2007gs} using the A14 set of the tuned parameters (tune) for the underlying event \cite{A14_tune} for the $W'$ and $\Graviton$ samples and the AZNLO tune \cite{AZNLO_tune} for the $H$ samples. Potential signal and background interference effects were ignored in the signal modelling.

MC simulated events are also used to model background processes. The main background sources are $Z$ and $W$ bosons produced in association with jets ($Z$+jets and $W$+jets),  with significant contributions from top quark production (both $t\bar{t}$ pair and single-top) and non-resonant vector-boson pair production ($ZZ$, $WZ$ and $WW$). 
The $Z$+jets and $W$+jets events were simulated using the \textsc{Sherpa} 2.2.1~\cite{Gleisberg:2008ta} event generator. Matrix elements were calculated for up to two partons at NLO and up to four partons at LO using the \textsc{Comix}~\cite{Gleisberg:2008fv} and \textsc{OpenLoops}~\cite{Cascioli:2011va} programs. Diboson processes with one of the bosons decaying hadronically and the other leptonically were simulated using \textsc{Sherpa} 2.1.0. They were simulated for up to one ($ZZ$) or zero ($WZ,\;WW$) additional partons at NLO and  up to three additional partons at LO using the \textsc{Comix} and \textsc{OpenLoops} programs. For both $Z$+jets and diboson simulation, the matrix-element calculations were merged with the \textsc{Sherpa} parton shower using the ME+PS@NLO prescription~\cite{Hoeche:2012yf}. The CT10 PDF set was used in conjunction with a dedicated parton shower tuning developed by the \textsc{Sherpa} authors. The $Z\to \tau\tau$ and $W\to \tau\nu$ events were included in the $Z$+jets and $W$+jets samples.
For the generation of top quark pairs, the \textsc{Powheg-Box} v2~\cite{Frixione:2007nw} event generator with the CT10 PDF set in the matrix element calculations was used. Electroweak $t$-channel , $s$-channel and $Wt$-channel single-top quark events were generated using the \textsc{Powheg-Box} v1 event generator~\cite{Alioli:2009je,Frederix:2012dh,Re:2010bp}.
This event generator uses the four-flavour scheme for the NLO matrix-element calculations together with the fixed four-flavour PDF set CT10f4~\cite{CT10}.  For all top quark processes, top quark spin correlations were preserved (for $t$-channel, top quarks were decayed using \textsc{MadSpin}). 
The parton shower, fragmentation, and underlying event were simulated using \textsc{Pythia}~6.428~\cite{Sjostrand:2006za} with the \textsc{CTEQ6L1}~\cite{1126-6708-2002-07-012} PDF set and the corresponding Perugia 2012 tune (P2012)~\cite{Skands:2010ak}.
The top quark mass was set to 172.5 GeV. The \textsc{EvtGen} v1.2.0 program~\cite{EvtGen} was used to decay bottom and charm hadrons for the \textsc{Powheg-Box} samples.

Cross sections were calculated with up to next-to-next-to-leading order (NNLO) QCD corrections for $Z$+jets and $W$+jets production~\cite{zxsec_NNLO}. Cross sections for diboson production were calculated at NLO including LO contributions with two additional partons~\cite{Gleisberg:2008ta,Hoche:2010kg}. The $t\bar{t}$ production cross section is calculated at NNLO in QCD, including resummation of next-to-next-to-leading logarithmic (NNLL) soft-gluon terms \cite{ttbarxsec1,ttbarxsec2}. The single-top production cross sections were calculated  to NLO in QCD \cite{topschxsec}, including the soft-gluon resummation at NNLL \cite{wtxsec} for the $Wt$ process.

In the simulation of the $Z$+jets events for the $ZV\to\llqq$ search, jets are labelled according to the true flavours of the hadrons with $\pt>5\,\gev$  found within a cone of size $\Delta R = \sqrt{(\Delta \eta)^2 + (\Delta \phi)^2} = 0.3$ around the reconstructed jet axes.
If a $b$-hadron is found, the jet is labelled as a $b$-jet;  if one is not found and a charmed hadron is found, the
jet is labelled as a $c$-jet; if neither is found, the jet is labelled as a light (i.e., $u$-, $d$-, or $s$-quark, or gluon) jet.
A simulated $Z$+jets event is labelled as $Z$+heavy-flavour if a $b$- or $c$-jet is found in the event, and otherwise as $Z$+light-flavour.
The classification allows for flavour-dependent corrections for differences between data and MC.

MC events were processed with a detailed detector simulation \cite{ATLAS_simu} based on \textsc{Geant4} \cite{Agostinelli:2002hh}. Additional inelastic simulated $pp$ collisions generated with {\textsc Pythia~8.186} were overlaid to model both the in- and out-of-time effects from additional $pp$ collisions in the same and neighbouring bunch crossings~(pile-up). MC samples were reweighted to match the pile-up conditions in the data.
All simulated events were processed using the same reconstruction algorithms as the data.

%% file: tex/objects.tex
Electrons are identified as isolated energy clusters in the electromagnetic calorimeter matched to
ID tracks, with requirements that the transverse energy $\et>7$~\GeV\ and pseudorapidity $|\eta|<2.47$. 
A likelihood-based requirement~\cite{ATLAS-CONF-2014-032} is imposed to reduce the background from misidentified or non-prompt electrons. Electrons are classified as either `loose', `medium' or `tight' according to the likelihood-based identification criteria described in Ref.~\cite{ATLAS-CONF-2014-032}.

Muons are reconstructed by a combined fit to the ID and MS tracks, and are required to have $\pt>7$~\GeV\ and $|\eta|<2.5$. Muons must pass identification requirements, based on the numbers of hits in the ID and MS subsystems, and the significance of the difference $|q/p_{\mathrm {MS}} - q/p_{\mathrm {ID}}|$~\cite{Aad:2016jkr}, where $q$ is the charge and $p_{\mathrm {MS}}\ (p_{\mathrm {ID}})$ is the momentum of the muon measured in the MS (ID). Similar to electrons, muons are classified as either `loose', `medium' or `tight', following the criteria in Ref.~\cite{Aad:2016jkr}.

All electrons and muons are required to be isolated using selections on the sum of track $\pt$ (excluding the track associated with the lepton) in a $\pT$-dependent cone around their directions. The isolation selection criteria are designed to maintain a constant efficiency of 99\% in the $\pt$--$\eta$ plane for the reconstructed leptons from $Z\to\ell\ell$ decays, and to minimise efficiency loss for highly boosted $Z$ bosons in the relevant kinematic range. 
Furthermore, leptons are required to have associated tracks satisfying $|d_0/\sigma_{d_0}|<5~(3)$ and $|z_0\times\sin\theta|<0.5$ mm for electrons~(muons), where $d_0$ is the transverse impact parameter with respect to the beam line, $\sigma_{d_0}$ is its uncertainty, and $z_0$ is the distance between the longitudinal position of the track along the beam line at the point where $d_0$ is measured and the longitudinal position of the primary vertex.

Small-$R$ jets are reconstructed using the anti-$k_t$ algorithm~\cite{antikt_algorithm,Cacciari:2011ma} with a radius parameter of $R=0.4$. Energy- and $\eta$-dependent correction factors derived from MC simulations are applied to correct jets back to the particle level~\cite{Aaboud:2017jcu}. Jets are required to have $\pT>20$~GeV for $|\eta|<2.5$ and $\pT>30$~GeV for $2.5<|\eta|<4.5$. 
Jets with $\pT>30$~GeV and $|\eta|<2.5$ are called `signal' jets. To suppress jets from pile-up interactions, a jet vertex tagger~\cite{JVT_Run1_perf} is applied to jets with $\pT<60$~GeV and $|\eta|<2.4$, based on information about tracks associated with the primary vertex and pile-up vertices.

The `signal' jets containing $b$-hadrons are identified using a multivariate algorithm ($b$-tagging)~\cite{Aad:2015ydr} which is based on information such as track impact-parameter significances and positions of explicitly reconstructed secondary decay vertices. The $b$-tagging is used for identifying $Z\to bb$ decays.
The chosen $b$-tagging algorithm has an efficiency of 70\% for $b$-quark jets in simulated $t\bar{t}$ events, with a light-flavour jet rejection factor of about 380 and a $c$-jet rejection of about 12~\cite{ATL-PHYS-PUB-2016-012}. Correction factors are applied to the simulated event samples to compensate for differences between data and simulation in the $b$-tagging efficiency for $b$-, $c$- and light-jets. The correction for $b$-jets is derived from $t\bar{t}$ events with final states containing two leptons, and the corrections are consistent with unity with uncertainties at the level of a few percent over most of the jet $\pT$ range.

Large-$R$ jets are reconstructed with the anti-$k_t$ algorithm, but with the radius parameter increased to $R=1.0$. 
To mitigate the effects of pile-up and soft radiation, the large-$R$ jets are trimmed~\cite{jet_trimming}.
Trimming takes the original constituents of the jet and reclusters them using the $k_t$ algorithm \cite{kt_algorithm} with a smaller radius parameter, $R_{\mathrm {subjet}}$, to produce a collection of subjets. These subjets are discarded if they carry less than a specific fraction ($f_{\mathrm{cut}}$) of the original jet \pT. The trimming parameters optimised for this search are $R_{\mathrm {subjet}} =0.2 $ and $f_{\mathrm{cut}}=5\%$. 
The large-$R$ jet four-momenta are recomputed from the selected subjets, and the jet energies are calibrated to particle-level using correction factors derived from MC simulations~\cite{Aad:2013gja}. 
The mass of a large-$R$ jet~($m_J$) is computed using a combination of calorimeter and tracking information~\cite{CombMass}. The large-$R$ jets are required to have $\pt > 200$~GeV and $|\eta|<2.0$. 

A default ATLAS-wide overlap-removal procedure is applied to the selected leptons and jets to prevent double-counting. 
For nearby electrons and small-$R$ jets, the jet is removed if the separation between the electron and jet is within
$\Delta R<$ 0.2; the electron is removed if the separation is within 0.2 $<\Delta R<$ 0.4.
For nearby muons and small-$R$ jets, the jet is removed if the separation between the muon and jet is within
$\Delta R<$ 0.2 and if the jet has less than three tracks or the energy and momentum differences between the muon and the jet are small; otherwise the muon is removed if the separation satisfies $\Delta R<$ 0.4.
To prevent double-counting of energy from an electron inside the large-$R$ jet, the large-$R$ jet is removed if the separation between the electron and the large-$R$ jet is within $\Delta R < 1.0$.

Boson tagging~\cite{Aad:2015rpa,ATL-PHYS-PUB-2015-033} is applied to the large-$R$ jets to select those from the $V\to qq$ decays.
A $\pT$-dependent requirement is applied to the jet substructure variable $\DTwoBetaOne$,
which is defined as a ratio of two- and three-point energy correlation functions~\cite{Larkoski:2014gra,Larkoski:2015kga} that are based on the energies and pairwise angular distances of particles within a jet. This variable is optimised to distinguish between jets originating from a single parton and those coming from the two-body decay of a heavy particle. 
A detailed description of the optimisation can be found in Refs.~\cite{Aad:2015rpa,ATL-PHYS-PUB-2015-033}.
The $V$ boson jets are then selected by requiring the large-$R$ jet mass $m_J$ to be in a $\pT$-dependent window  centred around the expected value of the boson mass from simulations. The boson tagging is designed to provide a constant efficiency (working point)
independent of the large-$R$ jet $\pT$ for the signals studied. Two such working points, 50\% efficiency and 80\% efficiency, are
used, with corresponding misidentification rates for jets from multijet production of $\sim 2$\% and $\sim 10$\%, respectively. 
For the 50\% working point, the width of the $W\, (Z)$ mass window varies from 23~(28) GeV at $\pT = 500$~GeV to 33 (37) GeV at $\pT = 2000$~GeV;
the maximum $D_2$ value for the $W\,(Z)$ tagger varies from 1.25 (1.22) at $\pT = 500$~GeV to 1.97 (1.85) at $\pT \ge 2000$~GeV.
For the 80\% working point, the width of the $W\, (Z)$ mass window varies from 33 (42) GeV at $\pT = 500$ GeV to 54 (57) GeV at $\pT = 2000$ GeV;
the maximum $D_2$ value for the $W\,(Z)$ tagger varies from 1.92 (1.90) at $\pT = 500$ GeV to 2.76 (2.62) at $\pT \ge 2000$ GeV.

The missing transverse momentum (\metvec) is calculated as the negative vectorial sum of the transverse momenta of calibrated electrons, muons, and small-$R$ jets. Large-$R$ jets are not included in the \metvec\ calculation to avoid double-counting of energy between the small-$R$ jets and large-$R$ jets. Energy depositions due to the underlying event and other soft radiation are taken into account by constructing a `soft term' from ID tracks associated with the primary vertex but not with any reconstructed object~\cite{Aad:2016nrq,ATL-PHYS-PUB-2015-027,ATL-PHYS-PUB-2015-023}. 
The track-based missing transverse momentum, $\mptvec$, is the negative vectorial sum of the transverse momenta of all good-quality inner detector tracks that are associated with the primary vertex.

%% file: tex/llqq.tex
\newcommand{\hsp}{\hspace*{0.3cm}}

\begin{figure}[h!]
\centering
  \includegraphics[width=0.85\textwidth]{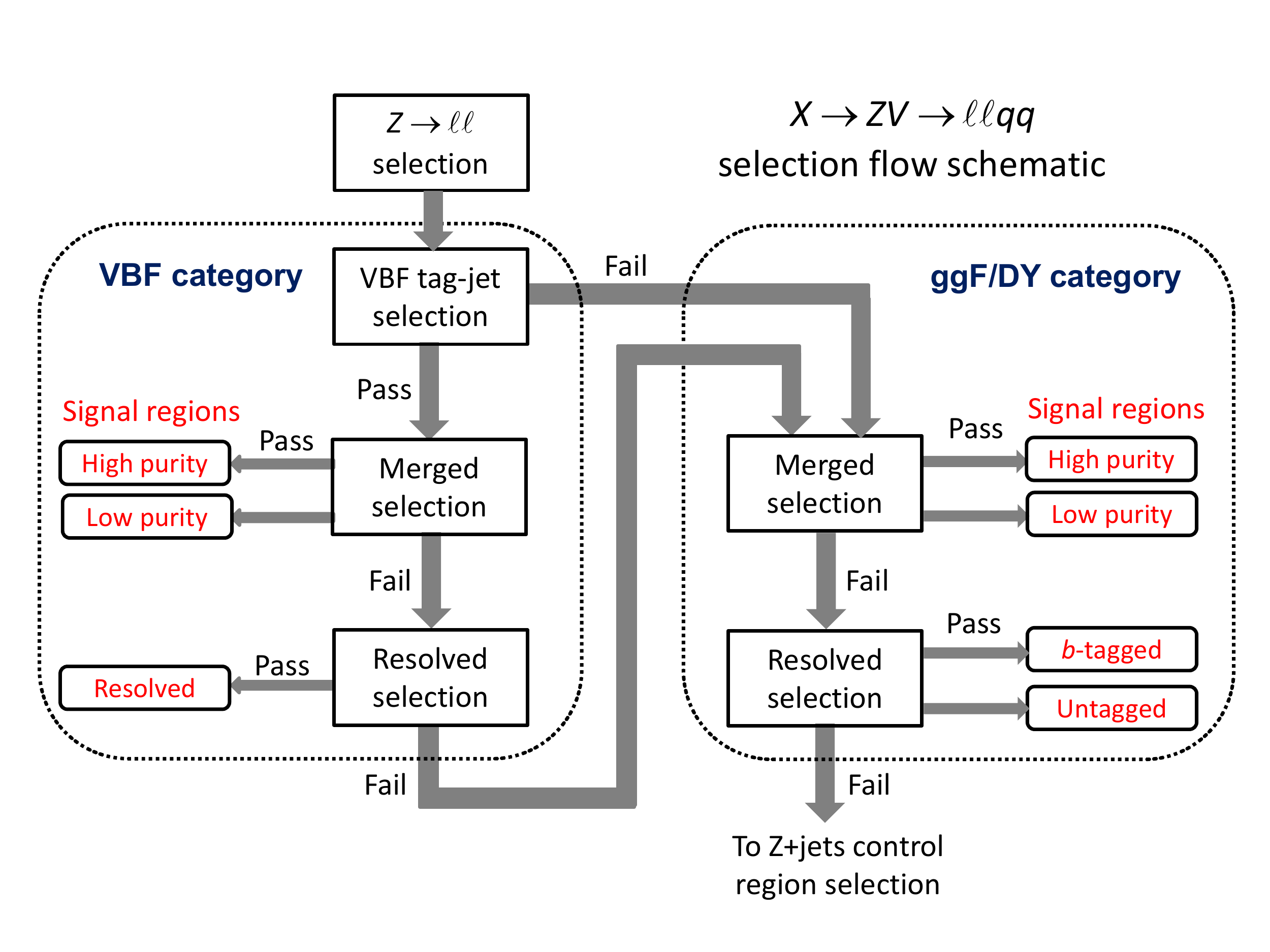}
  \caption{\label{fig:llqq_schematic} Illustration of the selection flow and seven signal regions of the $X\to ZV\to\llqq$ search.  The VBF category is targeted for VBF production whereas the ggF/DY category is for the rest, including events failing the selection for the VBF category. The selected VBF tag-jets are removed from the subsequent selection for the VBF category. However, if an event fails to be selected for the VBF category, these jets are kept for the ggF/DY category selection. The $H\to ZZ$ search utilises all seven signal regions and the $W'\to ZW$ search uses six signal regions by combining the $b$-tagged and untagged regions of the ggF/DY category. The $\Graviton\to ZZ$ search bypasses the VBF selection, so it has only four signal regions.
}
\end{figure}

The $X\to ZV\to\llqq$ search explores the VBF and ggF production of a heavy Higgs boson $H$,  the VBF and DY production of an HVT $W'$ boson, and the ggF production of a bulk RS graviton $\Graviton$. It also utilises both the merged and resolved reconstruction for the $V\to qq$ decay. The search begins with the identification of the $Z\to\ell\ell$ decay, followed by classifying events into the VBF or ggF/DY categories and finally the selection of either the $ZV\to\llJ$ or $ZV\to\lljj$ final states. Multiple signal regions~(SRs) are defined to enhance the sensitivity of the search. Figure~\ref{fig:llqq_schematic} shows a schematic view of the selection. The event selection, the expected signal performance and background estimations are described below.

\subsection{Selection of $Z\to\ell\ell$}
\label{sec:llqq_selection}

The $Z\to\ell\ell$ candidates are identified by requiring two isolated same-flavour leptons (electrons or muons) satisfying the `loose' criteria. The leading electron~(muon) must satisfy $\et~(\pt) > 28$~GeV. Opposite charges are required for the muon pairs but not for the electron pairs.  Electrons are more susceptible to charge misidentification due to the conversions of photons from bremsstrahlung, especially at high \et. The dilepton invariant mass $m_{\ell\ell}$ is required to be consistent with the $Z$ boson mass. For electrons, a fixed $m_{\ell\ell}$ window $[83,\, 99]$~GeV is applied. To account for the effect of  dimuon mass-resolution degradation at high  transverse momentum~($\ptll$), a $\ptll$-dependent mass window $[85.6~{\mathrm {GeV}}-0.0117\ptll,\, 94.0~{\mathrm {GeV}}+0.0185\ptll]$ is used for muons.  For both electrons and muons, the mass windows are chosen to ensure that the $Z\to\ell\ell$ selection efficiency is approximately 95\% and independent of the heavy resonance mass.  Events with additional leptons passing the `loose' criteria are vetoed.

\subsection{VBF and ggF/DY categories}
\label{sec:llqq_VBF}

Signal events from VBF production possess unique kinematic signatures. In addition to the presence of a pair of vector bosons from the resonance decay, VBF events have two additional jets (referred to as tag-jets). These jets typically have a large separation in pseudorapidity and a large dijet invariant mass. They offer a powerful means for background rejection. Hence for the $H\to ZZ$ and $W'\to ZW$ searches, for which the VBF production is considered, two categories are introduced. Events are subject to the selection designed for VBF production first (VBF category), and if they fail, to the selection designed for ggF or DY production (ggF/DY category). For the $\Graviton\to ZZ$ search, the VBF selection is bypassed as VBF production is not considered.

For the VBF category, events are required to have two tag-jets identified from the small-$R$ jets that fail the $b$-tagging described in Section~\ref{sec:objects}. The two jets must be in opposite pseudorapidity hemispheres,  i.e., $\eta_1\cdot\eta_2<0$, with a pseudorapidity separation $|\Delta\eta^{\mathrm {tag}}_{jj}|>4.7$, and  have an invariant mass $m^{\mathrm {tag}}_{jj}>770$~GeV. Those values are chosen to optimise the search sensitivity to VBF signals for all masses considered. If there is more than one pair of jets satisfying these requirements, the one with the highest $m^{\mathrm {tag}}_{jj}$ value is chosen. These jets are not considered in the $ZV\to\llqq$ selection, and large-$R$ jets lying $\Delta R<1.5$ of either of  these two small-$R$ jets are not considered either.

For the ggF/DY category, no tag-jets requirement is applied. Events in the VBF and ggF/DY categories are subject to an identical $ZV\to\ell\ell qq$ selection.  All events not selected for the VBF category are passed to the ggF/DY selection.
This includes events that contain jets satisfying the VBF requirements but which fail the $ZV\to\llqq$ selection, cf.~Figure~\ref{fig:llqq_schematic}. In this case, jets excluded from consideration for the $ZV\to\llqq$ candidate in the VBF category are considered for the ggF/DY category.

\subsection{Selection of $ZV\to\llqq$}

Identification of $ZV\to\llqq$ decays proceeds by applying the merged $ZV\to\llJ$ selection followed by the resolved $ZV\to\lljj$ selection. The order is motivated by a smaller background expectation in the $ZV\to\llJ$ final states. These selection criteria are summarised in Table~\ref{tab:llqq_selection} and are explained below.

\subsubsection{Merged $ZV\to\llJ$ selection}
\label{sec:llqq_merged}
The $ZV\to\llJ$ candidates are selected from the $Z\to\ell\ell$ events containing at least one large-$R$ jet, among which the one with the highest \pT\ is assumed to be from the $V\to qq$ decay.
Events are further required to have  $\min(\ptll,\ptJ)/\mllJ>0.3$ for the $H\to ZZ$ search and $>0.35$ for the $W'\to ZW$ and $\Graviton\to ZZ$ searches. The criteria are optimised to reduce backgrounds while retaining high efficiencies for signals. The looser requirement for the $H\to ZZ$ search is motivated by the expected softer $\ptll$ and $\ptJ$ spectra resulting from a spin-0 resonance. This requirement suppresses background significantly at large $\mllJ$ while maintaining high efficiencies for signal events.

\begin{figure}[h!]
\centering
  \subfigure[]{\includegraphics[width=0.495\textwidth]{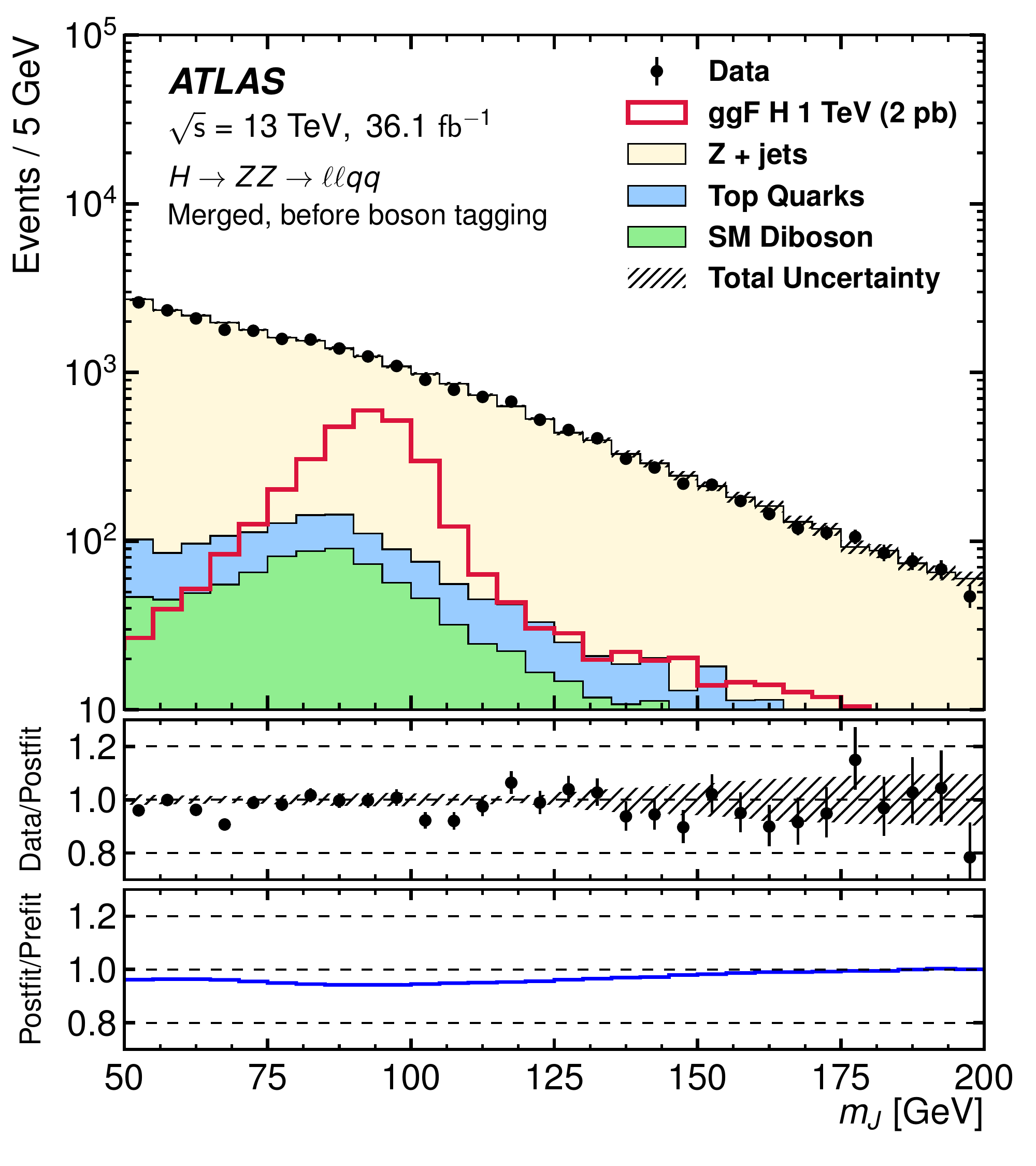}}
  \subfigure[]{\includegraphics[width=0.495\textwidth]{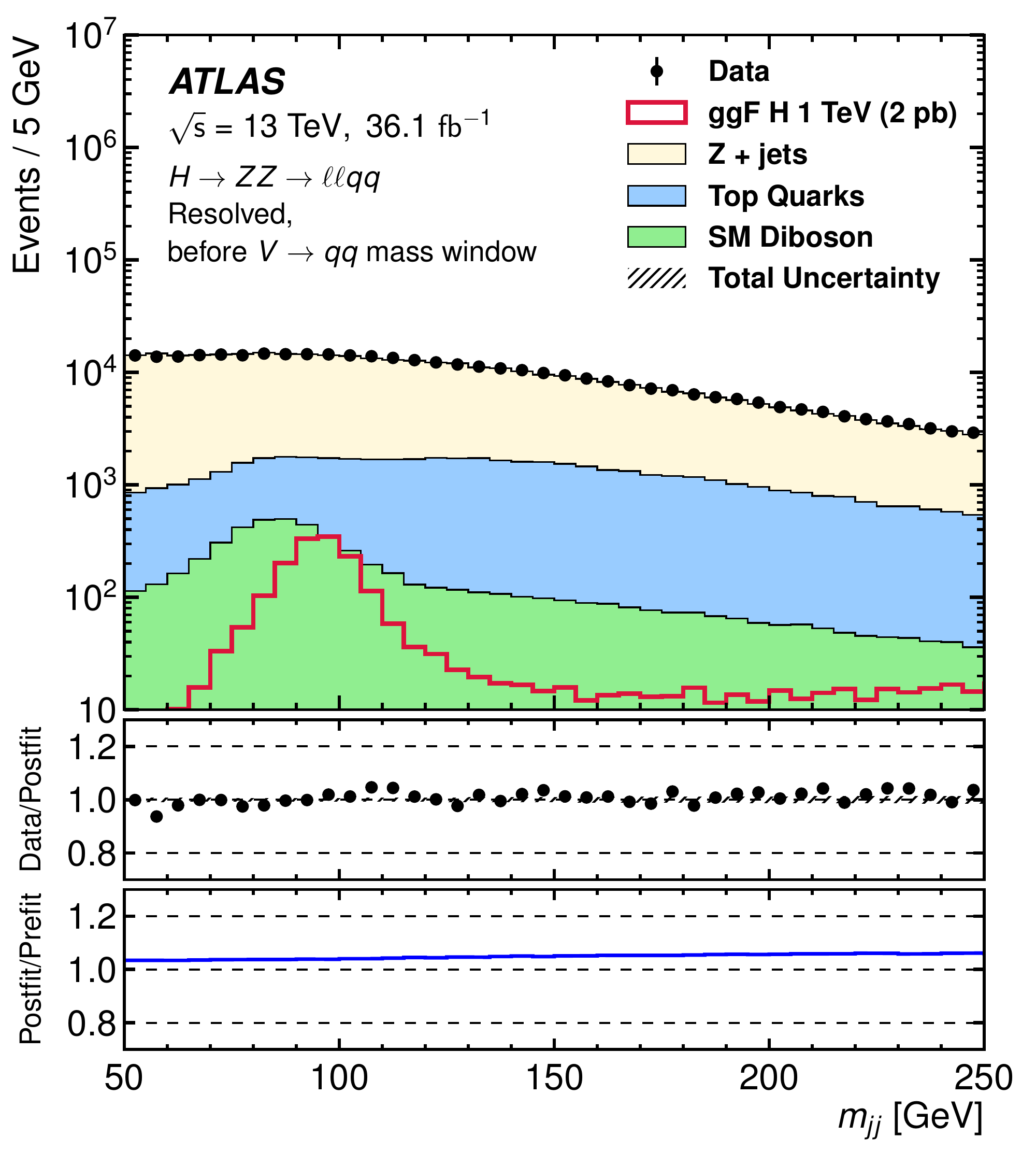}}
  \caption{\label{fig:llqq_mjets} Observed and expected distributions of (a) the large-$R$ jet mass $m_J$ of the $ZV\to\llJ$ candidate and (b) the dijet mass $m_{jj}$ of the $ZV\to\lljj$ candidate. These distributions are for the $H\to ZZ$ search before the identification of the $V\to qq$ decay, combining VBF and ggF/DY categories. Background contributions are obtained from the combined likelihood fit to the data, as explained in Section~\ref{sec:results}. For illustration, expected distributions from the ggF production of a 1~TeV Higgs boson with $\sigma\times\mathcal{B}(H\to ZZ)=2$~pb are also shown.
The middle panes show the ratios of the observed data to the background predictions. The uncertainty in the total background prediction, shown as bands, combines statistical and systematic contributions. The bottom panes are the ratios of the  post-fit and pre-fit background predictions.}
\end{figure}

Figure~\ref{fig:llqq_mjets}(a) shows the distribution of the large-$R$ jet mass $m_J$ of the $H\to ZZ$ search, comparing data with the expected backgrounds. Next the boson tagging discussed in Section~\ref{sec:objects} is applied to select the $V\to qq$ decays. Two signal regions are defined, one for events passing the 50\% working point of the boson tagging requirement and the other for events failing the 50\% but passing the 80\% working point requirement. The former is called the high-purity~(HP) SR, and the latter the low-purity~(LP) SR.   Background contributions are mostly from the  $Z$+jets, top quark and diboson production (Section~\ref{sec:llqq_background}).

\begin{table}[htb]
\caption{Summary of the $X\to ZV\to \llqq$ selection criteria. }
\vspace*{0.2cm}
\begin{center}
  \begin{tabular}{clclclc}
  \toprule
 & Selection              &&    $ZV\to\llJ$      &\hsp&   $ZV\to\lljj$  &  \\ \hline
 & $Z\to\ell\ell$         && \multicolumn{3}{c}{Two opposite-flavour leptons with $\pT(\ET)>7$~GeV} \\
 &                        && \multicolumn{3}{c}{leading lepton with $\pT(\ET)>28$~GeV\hspace*{0.5cm}} \\
 &                        && \multicolumn{3}{c}{$83<m_{ee}<99$~GeV\hspace*{0.5cm}} \\
 &                        && \multicolumn{3}{c}{$85.6\,\gev -0.0117\times\ptll<m_{\mu\mu}<94.0\,\gev +0.0185\times\ptll$\hspace*{1.2cm}} \\ \\[-1mm]\cline{2-6}

\\[-4mm]
 & Tag-jet selection for  && \multicolumn{3}{c}{Two non-$b$-tagged  small-$R$ jets with } \\
 &  VBF category          && \multicolumn{3}{c}{$\eta_1\cdot\eta_2<0$, $\left|\Delta\eta_{jj}^{\mathrm {tag}}\right|>4.7$ and $m^{\mathrm {tag}}_{jj}>770$~GeV} \\ \\[-1mm]\cline{2-6}

\\[-4mm]
 & Jet requirements       && $\ge 1$ large-$R$ jet       &&  $\ge 2$ `signal' jets with $\pT>30$~GeV   \\
 &                        && with $\pT>200$~GeV          &&  $\pT>60$~GeV for the leading jet       \\
 &                        &&                             &&  no events with $>2\, b$-tagged jets  \\  \\[-1mm]\cline{2-6}

\\[-4mm]
 & Kinematic criteria      &&  $\min(\ptll,\ptJ)\left/\mllJ\right.$   && $\sqrt{\left(\ptll\right)^2+\left(\ptjj\right)^2}\left/\mlljj\right.$ \\
 & \hsp\hsp $H$            &&  \hsp\hsp $>0.3$        && \hsp\hsp $>0.4$   \\
 & \hsp\hsp $W'$ or $\Graviton$  &&  \hsp\hsp $>0.35$       && \hsp\hsp $>0.5$   \\ \\[-1mm]\cline{2-6}

\\[-4mm]
 & $V$ boson tagging       && $\pt$-dependent criteria         &&  $70<\mjj<105$~GeV ($V=Z$) \\
 &                         && in $\DTwoBetaOne$ and $m_J$   &&  $62<\mjj<97$~GeV ($V=W$) \\
  \bottomrule

  \end{tabular}
\end{center}
\label{tab:llqq_selection}
\end{table}

\subsubsection{Resolved $ZV\to\lljj$ selection}
\label{sec:llqq_resolved}
The $ZV\to \ell\ell jj$ candidates are selected from the $Z\to\ell\ell$ events that have failed the $ZV\to\llJ$ selection, by requiring at least two small-$R$ `signal' jets. The leading jet $\pt$ is required to be greater than 60~GeV.

Similarly to the merged $ZV\to\llJ$ selection, the kinematic quantity  $\sqrt{\left(\ptll\right)^2+\left(\ptjj\right)^2}\left/m_{\ell\ell jj}\right.$ is required to be greater than $0.4$ for $H\to ZZ$ and $0.5$ for $W'\to ZW$ and $\Graviton\to ZZ$. Here $\ptjj$ is the transverse momentum of the dijet system.
Signal events are expected to have a dijet invariant mass, $m_{jj}$, consistent with a $V\to qq$ decay. Figure~\ref{fig:llqq_mjets}(b) shows the $m_{jj}$ distributions of the $H\to ZZ$ search. The dijet invariant mass must be in the window $[70,\, 105]$~GeV for $Z\to qq$  and in the window $[62,\, 97]$~GeV for $W\to qq$. The choice of asymmetric windows around $m_V$ is motivated by the asymmetry of the $m_{jj}$ distribution expected from $V\to qq$.

About 21\% of the $Z\to qq$ decays have two $b$-quark jets, whereas the dominant background, $Z+$jets, has a smaller heavy-quark content. To exploit this, the $ZV\to\lljj$ candidates are divided into two signal regions: events with two $b$-tagged jets ($b$-tagged SR) and events with fewer than two $b$-tagged jets (untagged SR). Events with more than two $b$-tagged jets are rejected.
No enhancement of $b$-tagged jets is expected from the $W\to qq$ decay, thus the two signal regions are combined for the $W'\to ZW$ search. Due to the small number of events, the two regions are also combined for the VBF category.

\subsection{Signal regions, selection efficiencies and mass resolutions}

Signal regions used in the search depend on the signal model under study. There are seven, six and four signal regions, respectively, for the searches for $H\to ZZ$, $W'\to ZW$ and $\Graviton\to ZZ$.

\begin{figure}[h!]
\centering
  \subfigure[]{\includegraphics[width=0.495\textwidth]{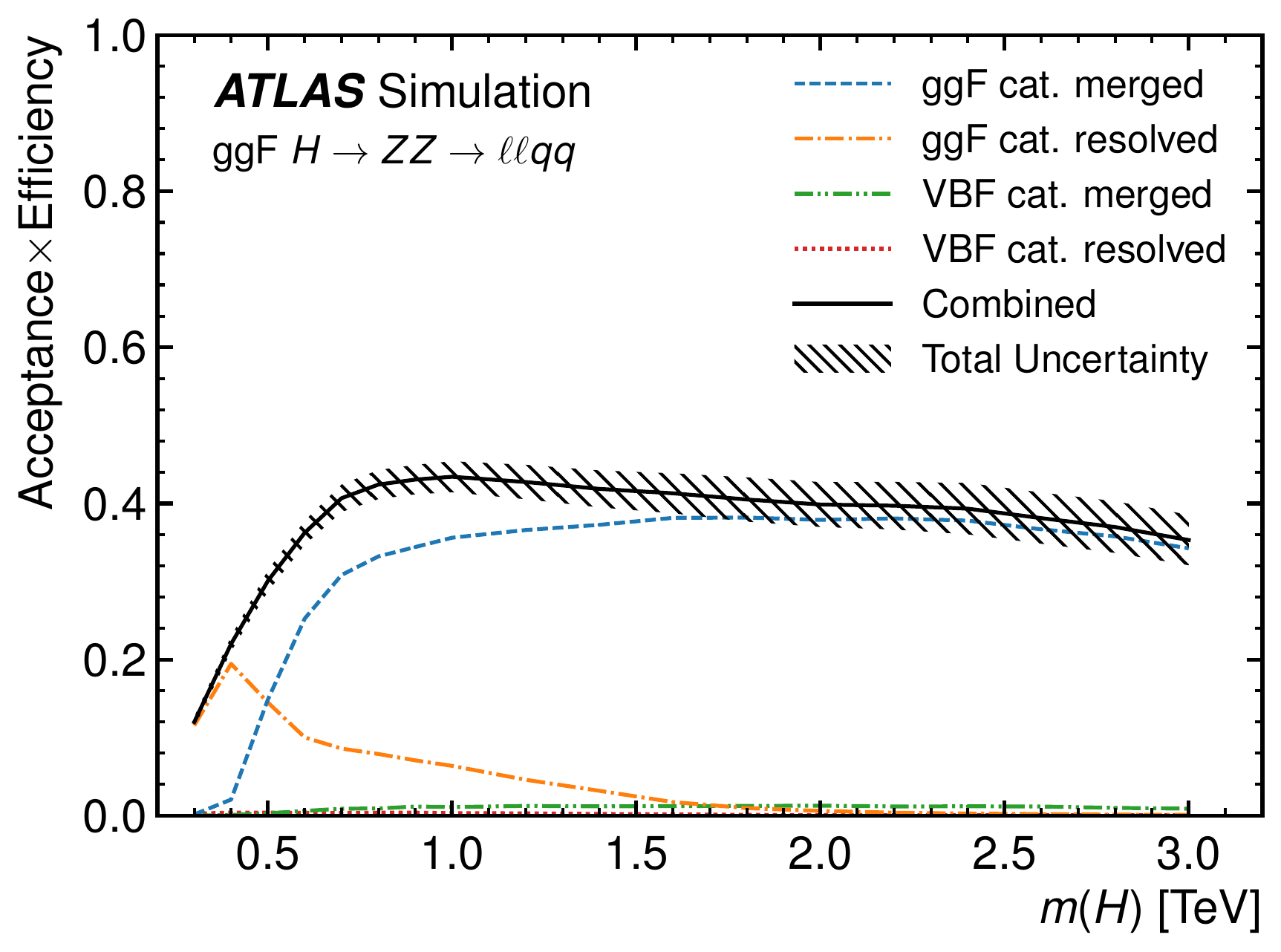}}
  \subfigure[]{\includegraphics[width=0.495\textwidth]{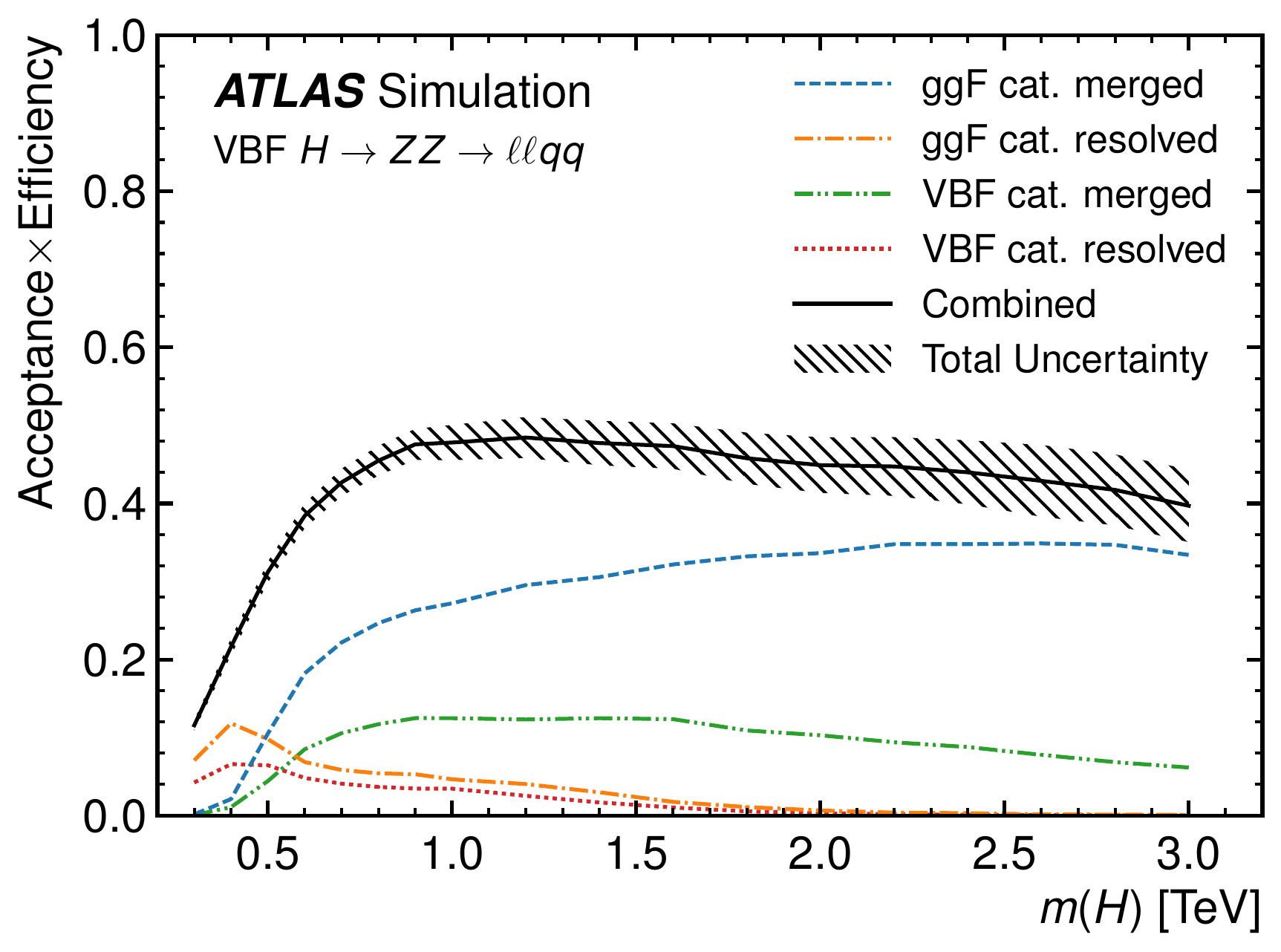}}
  \caption{\label{fig:llqq_efficiency} Selection acceptance times efficiency for the $H\to ZZ\to\llqq$ events from MC simulations as a function of the Higgs boson mass for (a) ggF and (b) VBF production, combining the HP and LP signal regions of the $ZV\to\llJ$ selection and the $b$-tagged and untagged regions of the  $ZV\to\lljj$ selection. The hatched band represents the total statistical and systematic uncertainties.}
\end{figure}

Signal selection efficiencies are dependent on the signal model, the production process and the mass of the heavy resonance.  As an example,  Figure~\ref{fig:llqq_efficiency} shows the acceptance times efficiency ($A\times\epsilon$) of the $H\to ZZ\to \llqq$ search as a  function of the Higgs boson mass for both ggF and VBF production. The $ZV\to\llJ$ selection is more efficient for masses over approximately 500~GeV, while the $ZV\to\lljj$ selection contributes more at lower masses.
This reflects the expected large Lorentz boost of the $Z$ bosons from heavy Higgs boson decays and the higher priority given to the $ZV\to\llJ$ selection. The $A\times\epsilon$ values of the $W'\to ZW$ and $\Graviton \to ZZ$ searches are similar with the exception of noticeable decreases, by about 40\% from 2~TeV to 4~TeV, due to the merging of the electrons from the $Z\to ee$ decays for resonance masses above approximately 3~TeV.

Distributions of the invariant masses $\mllJ$ and $\mlljj$ are used to search for potential signals. To mitigate the impact of  muon momentum-resolution degradation at high $\pT$,  a scale factor of $m_Z/m_{\mu\mu}$ is applied  to the four-momentum of the dimuon system in $Z\to\mu\mu$ events, effectively fixing its mass to $m_Z=91.187$~GeV~\cite{Agashe:2014kda}.
The scaling improves the $\mllJ$ resolution of the dimuon final state by approximately 13\%~(40\%) for a heavy Higgs boson at 1~(3)~TeV.
For the $ZV\to \lljj$ final states, a scale factor of $m_V/m_{jj}$ is also applied to the dijet four-momentum to fix the dijet mass to $m_V$, 91.187~GeV for the $Z$ boson and 80.385~GeV for the $W$ boson~\cite{Agashe:2014kda}.
The $m_{jj}$ scaling improves the $\mlljj$ resolution by approximately 14\% for a 600~GeV Higgs boson. No $m_V/m_J$ scale factor is applied to the $ZV\to\llJ$ final states as the expected improvement in the $m_{\llJ}$ resolution is negligible.

The reconstructed $\mllJ$ and $\mlljj$ distributions have widths of 2--3\% of the resonance mass for the Higgs boson and HVT $W'$ boson of VBF production for the entire mass range studied. Given the negligible intrinsic widths of these resonances, the widths largely reflect the detector resolution. For DY-produced HVT $W'$ signal, the widths of the $\mllJ$ and $\mlljj$ distributions are slightly larger at 3--4\% since the $W'$ boson in this production mode has an intrinsic width of approximately 2.6\% of its mass. The widths of the $\mllJ$ and $\mlljj$ distributions are 4\% at 500 GeV for the bulk RS graviton signal with $\kMPl=1$ and rise to 8\% at 5000~GeV. The increase can be attributed to the increase in the intrinsic width of the signal. For a given resonance mass, the $\mllJ$ distributions are narrower than those of $\mlljj$.

\subsection{Data control regions and background estimation}
\label{sec:llqq_background}

The dominant backgrounds to the $X\to ZV\to\llqq$ search are the $Z$+jets, top quark and diboson processes. Their contributions are estimated from a combination of MC and data-driven techniques. In all cases, the shapes of kinematic variables, including those of the final discriminants $\mllJ$ and $\mlljj$,  are taken from MC simulations. The multijet background is estimated to be negligible.

The $Z$+jets events are expected to have smooth distributions of $m_J$ and $m_{jj}$, while the signal events should exhibit resonance structures at the mass of the vector-boson $V$. Thus, a $Z$+jets control region~(CR) is defined for every signal region by reversing the $m_J$ or $m_{jj}$ requirement. Events in the control regions are selected in exactly the same way as those in their corresponding signal regions except for the requirement on $m_J$ or $m_{jj}$. For the $ZV\to\llJ$ selection, the leading large-$R$ jet mass is required to be outside the large-$R$ jet mass window of the 80\% working point of the boson tagging. For the $ZV\to\lljj$ selection, a requirement of $50<m_{jj}<62$~GeV or $105<m_{jj}<150$~GeV is applied.
These CRs are dominated by the $Z$+jets contribution, with a purity higher than $96$\% in all regions, except for the $b$-tagged CR where the top quark and $Z$+jets contributions are comparable. They are therefore used to constrain its contribution in signal regions through simultaneous fits as discussed in Section~\ref{sec:results}.

Top quark production is a significant background source in the $b$-tagged signal region of the resolved $ZV\to\lljj$ selection. Its contribution is constrained using a top-quark-enhanced control region. Events in this control region must have two different-flavour leptons, $e\mu$, with their invariant mass within $[76,\, 106]$~GeV, and two $b$-tagged jets with their invariant mass in the range $[50,\, 150]$~GeV. The leading $b$-tagged jet is required to have $\pT>60$~GeV. This selection yields a sample of top quark events with a purity higher than $99\%$. This top quark control region is used to constrain top quark contributions for all signal regions.

Diboson production, mainly from the SM $ZZ$ and $ZW$ processes, is also a significant background source.
The contribution from those processes is estimated completely from MC simulation.

As an example, Table~\ref{tab:llqq_yields} shows the numbers of events observed in the data and estimated from background processes in the seven signal regions of the $H\to ZZ\to \ell\ell qq$ search. The numbers of background events are extracted through a background-only fit of the signal and control regions discussed in Section~\ref{sec:results}.

\begin{table}[!hbt]
  \caption{\label{tab:llqq_yields} Numbers of events observed in the data and predicted for background processes from background-only fits to the signal and control regions~(Section~\ref{sec:results}) in the seven signal regions of the $H\to ZZ\to \llqq$ search. The numbers of signal events expected from the ggF and VBF production of a heavy Higgs boson with mass of 1~TeV are also shown. The signal yields are calculated using $\sigma\times\mathcal{B}(H\to ZZ)= 20$~fb for both processes. The uncertainties combine statistical and systematic contributions.
The fit constrains the background estimate towards the observed data, which reduces the total background uncertainty by correlating those from the individual backgrounds.
  }
  \centering
  \vspace*{0.2cm}
  \footnotesize
  \begin{tabular}{ll r@{\,\( \pm \)\,}l r@{\,\( \pm \)\,}l r@{\,\( \pm \)\,}l r@{\,\( \pm \)\,}l r@{\,\( \pm \)\,}l r@{\,\( \pm \)\,}l r}
  \toprule
    $V\to qq$  & Signal &\multicolumn{4}{c}{$H$ (1 TeV)}   & \multicolumn{8}{c}{Background estimates}  & \multirow{2}{*}{Data} \\ \cmidrule(lr){3-6} \cmidrule(lr){7-14}
    recon.     & regions  & \multicolumn{2}{c}{ggF} & \multicolumn{2}{c}{VBF}  & \multicolumn{2}{c}{$Z+$jets} & \multicolumn{2}{c}{Diboson} & \multicolumn{2}{c}{Top quarks}  & \multicolumn{2}{c}{Total} &  \\ \hline \\
    \multicolumn{15}{c}{VBF category} \\ \hline
    Merged     & HP          & $0.42$&$0.08$ & $5.1$&$1.0$   &  $29.0$&$2.6$  & $3.8$&$0.6$  & $1.1$&$0.4$   & $33.9$&$2.7$  & 32     \\
               & LP          & $0.33$&$0.08$ & $3.4$&$0.4$   &  $113$&$7$     & $8.4$&$1.2$  & $1.8$&$0.6$   & $123$&$7$     & 109    \\[2mm]
    Resolved   &             & $0.23$&$0.05$ & $2.3$&$0.4$   &  $1307$&$34$   & $60$&$9$     & $66$&$7$      & $1433$&$34$   & 1434   \\ \hline \\

    \multicolumn{15}{c}{ggF/DY category} \\ \hline
    Merged     & HP          & $14.2$&$1.6$  & $11.0$&$2.1$  &  $1728$&$34$   & $177$&$21$   & $20.6$&$2.2$  & $1926$&$32$   & 1906   \\
               & LP          & $10.0$&$0.9$  & $7.5$&$0.8$   &  $6060$&$60$   & $285$&$31$   & $69$&$6$      & $6420$&$60$   & 6375   \\[2mm]
    Resolved   & $b$-tagged  & $1.02$&$0.12$ & $0.62$&$0.08$ &  $1740$&$40$   & $167$&$22$   & $908$&$24$    & $2810$&$40$   & 2843   \\
               & Untagged    & $3.31$&$0.34$ & $2.5$&$0.5$   &  $82200$&$400$ & $2280$&$250$ & $1500$&$130$  & $86030$&$280$ & 85928  \\
  \bottomrule
  \end{tabular}
\end{table}

\clearpage

%% file: tex/vvqq.tex
Events containing a $X\to ZV\to\vvqq$ signal are characterised by a hadronically decaying $V$ boson recoiling against a large missing transverse momentum. Only the merged reconstruction of the $V\to qq$ decay is considered. The selection closely follows that of the $X\to ZV\to\llqq$ search as illustrated in Figure~\ref{fig:llqq_schematic} with additional requirements to remove multijet backgrounds but without the resolved selection.

\subsection{Selection of $ZV\to\vvqq$}
\label{sec:vvqq_selection}

An initial selection is made by requiring $\met >250$~ GeV,
and vetoing events with electrons or muons passing the `loose' quality requirements.
The multijet background, originating primarily from the presence of mismeasured jets, and non-collision backgrounds are suppressed by using a track-based missing transverse momentum, \mptvec, as defined in Section~\ref{sec:objects}.
The requirements are \mpt\ $>50$~GeV and the azimuthal separation between \metvec\ and \mptvec\ directions satisfies $\Delta \phi(\metvec, \mptvec)<1$.
An additional requirement is imposed on the azimuthal separation between the directions of \metvec\ and the nearest small-$R$ jet with $\min[\Delta\phi(\metvec,{\mathrm {small}}\mathrm{-}R~{\mathrm {jet}})]>0.4$. The multijet background is found to be negligible after these selections.

As in the \llqq\ search, both the VBF and ggF categories are considered in the signal event selection for the $H\to ZZ$ and $W'\to ZW$ searches; the VBF selection is not motivated by the $\Graviton\to ZZ$ search.
The VBF tag-jets are selected in the same manner as in the \llqq\ search, but they are required to satisfy
$m_{jj}^{\mathrm {tag}}>630$~GeV and $|\Delta\eta^{\mathrm {tag}}_{jj}|>4.7$. Those values are chosen to optimise the search sensitivity for signals from VBF production, considering all signal masses. Events failing the VBF tag-jets selection are kept in the ggF category.
For both the VBF and ggF categories, events are then subject to the selections described below.

\begin{figure}[hbt]
\centering
  \subfigure[]{\includegraphics[width=0.495\textwidth]{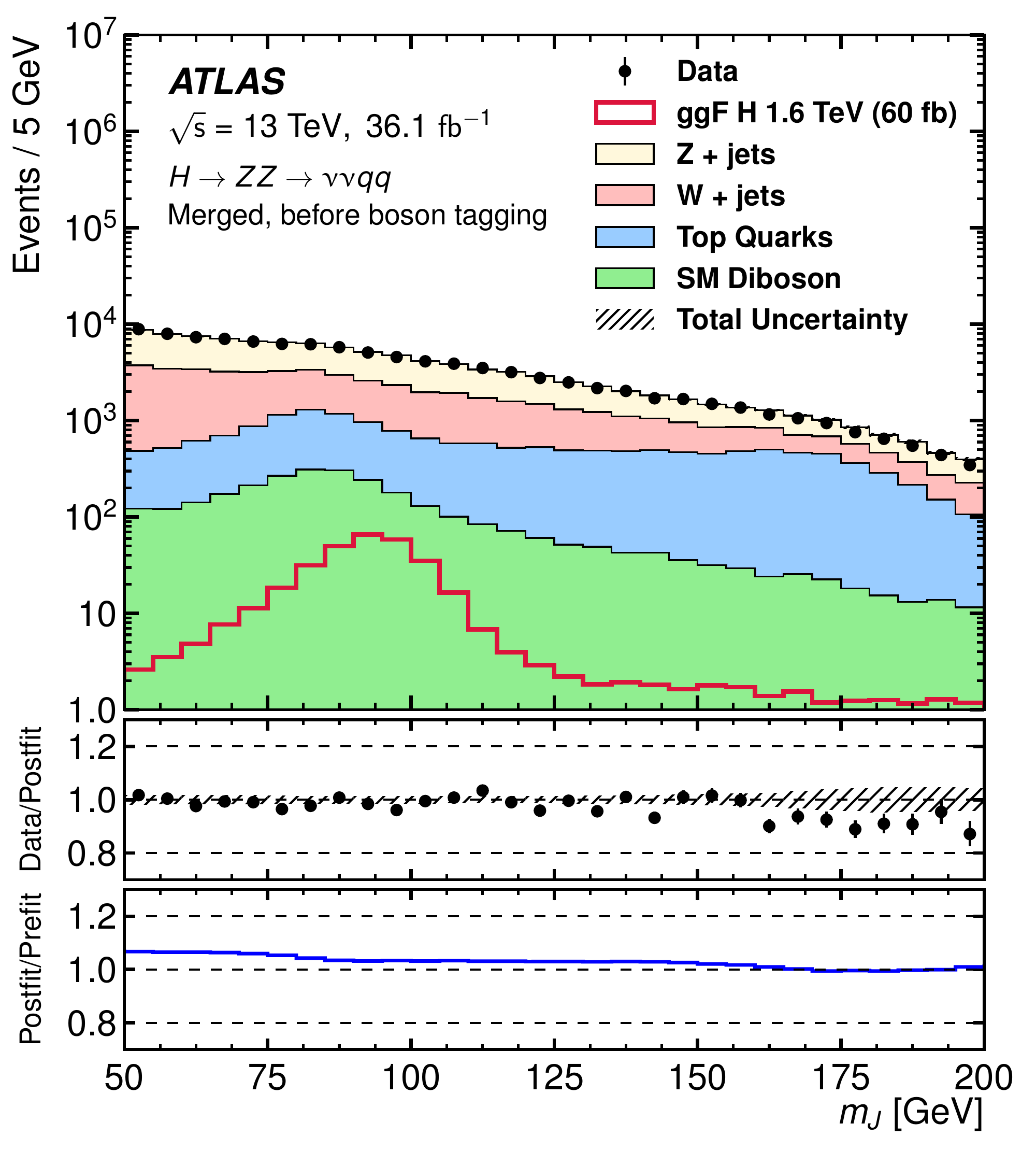}}
  \subfigure[]{\includegraphics[width=0.495\textwidth]{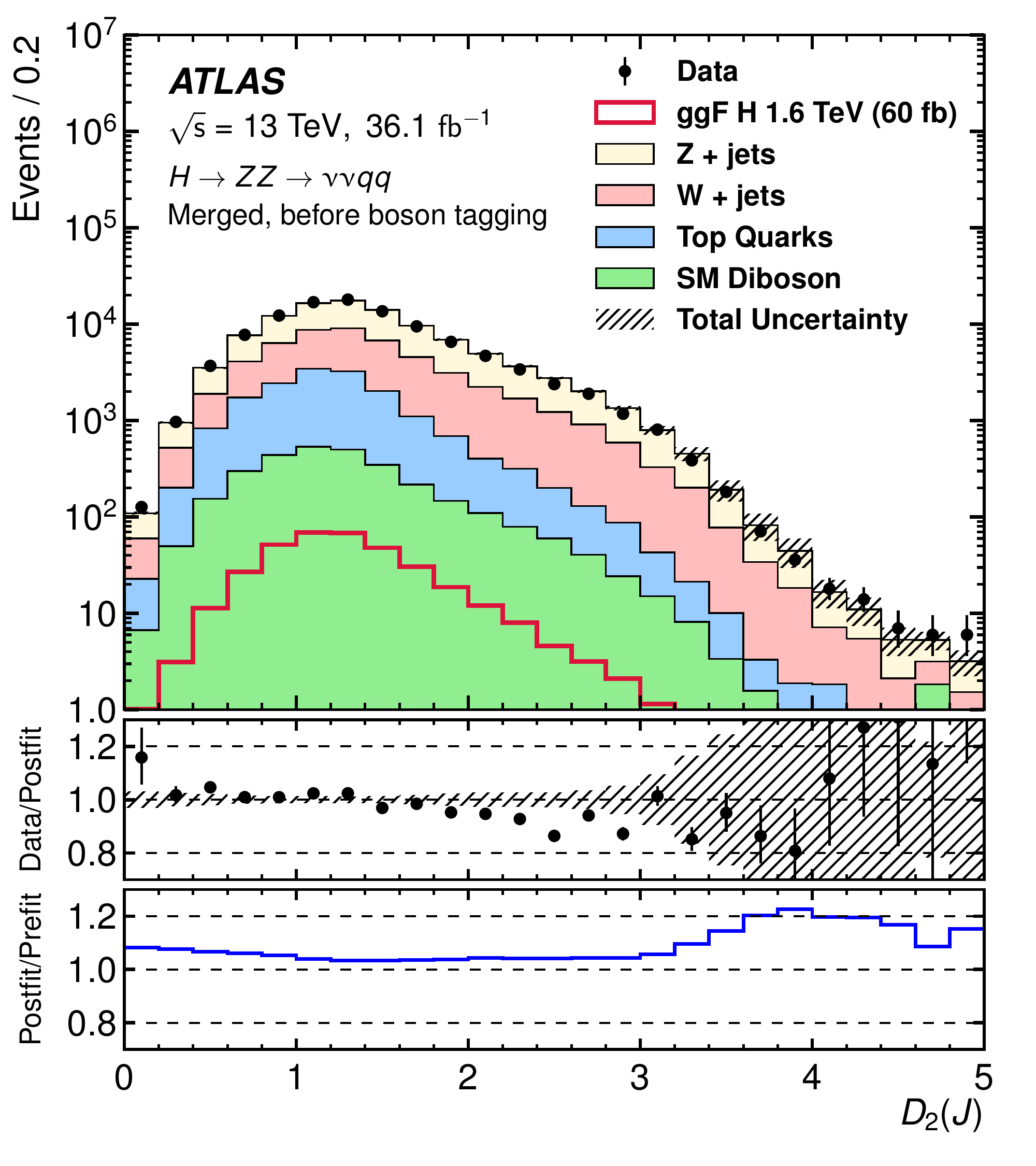}}

  \caption{\label{fig:vvqq_ptJ} Observed and expected distributions of (a) mass of the large-$R$ jet and (b) \DTwoBetaOne\  of  the large-$R$ jet in $ZV\to\vvJ$ candidates in the ggF category. All selections other than the requirements on the mass and \DTwoBetaOne\ are applied. Background contributions in these distributions are after applying the combined likelihood fit to the data, as explained in Section~\ref{sec:results}. For illustration, expected distributions from the ggF production of a 1.6~TeV Higgs boson with $\sigma\times\mathcal{B}(H\to ZZ)=60$~fb are also shown. The middle panes show the ratios of the observed data to the background predictions. The uncertainty in the total background prediction, shown as bands, combines statistical and systematic contributions. The bottom panes are the ratios of the post-fit and pre-fit background predictions. }
\end{figure}

Unlike the \llqq\ search, this search utilises only the merged selection of the $V\to qq$ decay, i.e., considering only the $ZV\to \nu\nu J$ final state, since the resolved selection has little acceptance for the given high \MET\ trigger threshold.  The selection of the large-$R$ jets and the boson tagging are identical to those of the \llqq\ search and are described in Section~\ref{sec:objects}. Similarly, candidate events are split into HP and LP signal regions.
 The $H\to ZZ$ search and the $W'\to ZW$ search use all the four signal regions, while the $\Graviton\to ZZ$ search bypasses the VBF selection, so has only two signal regions.
 The distributions of the mass and \DTwoBetaOne\ of the leading large-$R$ jet before the boson tagging are shown in Figure~\ref{fig:vvqq_ptJ}.
There is a small overprediction of the background at high mass region, but those events are not selected in any signal regions and sidebands.
The selection criteria for the \vvqq\ search are summarised in Table~\ref{tab:vvqq_selection}.

\begin{table}[!htb]
\caption{Summary of the $X\to ZV\to \vvqq$ selection criteria. }
\vspace*{0.2cm}
\begin{center}
  \begin{tabular}{lc}
  \toprule
 $Z\to\nu\nu$           &  $\met > 250$~GeV \\
 \hline
 \multirow{3}{*}{Multijet removal}      &  $\mpt > 50$~GeV  \\
                        &  $\Delta \phi(\metvec, \mptvec)<1$ \\
                        &  $\min[\Delta\phi(\metvec,{\mathrm {small}}\mathrm{-}R~{\mathrm {jet}})]>0.4$  \\
 \hline
 Tag-jet selection for  & Two non-$b$-tagged  small-$R$ jets with \\
  VBF category          & $\eta_1\cdot\eta_2<0$, $\left|\Delta\eta_{jj}^{\mathrm {tag}}\right|>4.7$ and $m^{\mathrm {tag}}_{jj}>630$~GeV \\
 \hline
 Jet requirements       & $\ge 1$ large-$R$ jet with $\pT>200$~GeV \\
 \hline
 $V$ boson tagging      & $\pt$-dependent criteria on $\DTwoBetaOne$ and $m_J$ \\
  \bottomrule
  \end{tabular}
\end{center}
\label{tab:vvqq_selection}
\end{table}

It is not possible to fully reconstruct the invariant mass of the \vvJ\ system due to the presence of neutrinos in the final state, so the transverse mass is used as the final discriminant:

\begin{equation}
\label{eq:mT}
m_{{\mathrm T}} = \sqrt{\left(E_{{\mathrm T},J} + E_{{\mathrm T}}^{\mathrm {miss}}\right)^2 - \left(\vec{p}_{{\mathrm T},J} + \vec{E}_{{\mathrm T}}^{\mathrm {miss}}\right)^2}, \nonumber
\end{equation}

where $E_{{\mathrm T},J} = \sqrt{m_{J}^2 + p_{{\mathrm T},J}^2}$.
The resolution of the transverse mass is about 10\% of the signal mass and the ratio of the resolution to the signal mass has little dependence on the signal model or the resonance mass.

\begin{figure}[!h!btp]
\centering
  \subfigure[]{\includegraphics[width=0.495\textwidth]{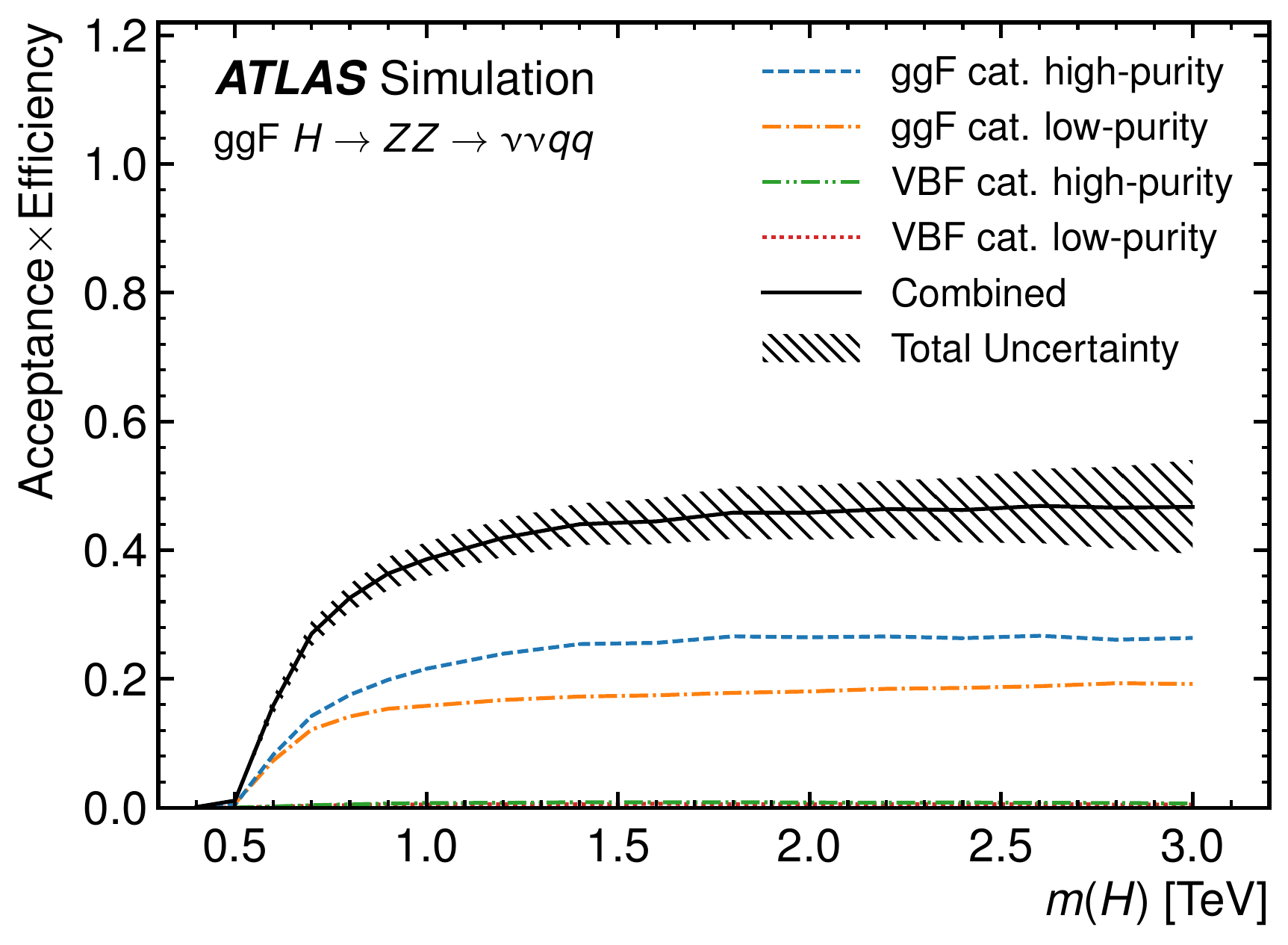}}
  \subfigure[]{\includegraphics[width=0.495\textwidth]{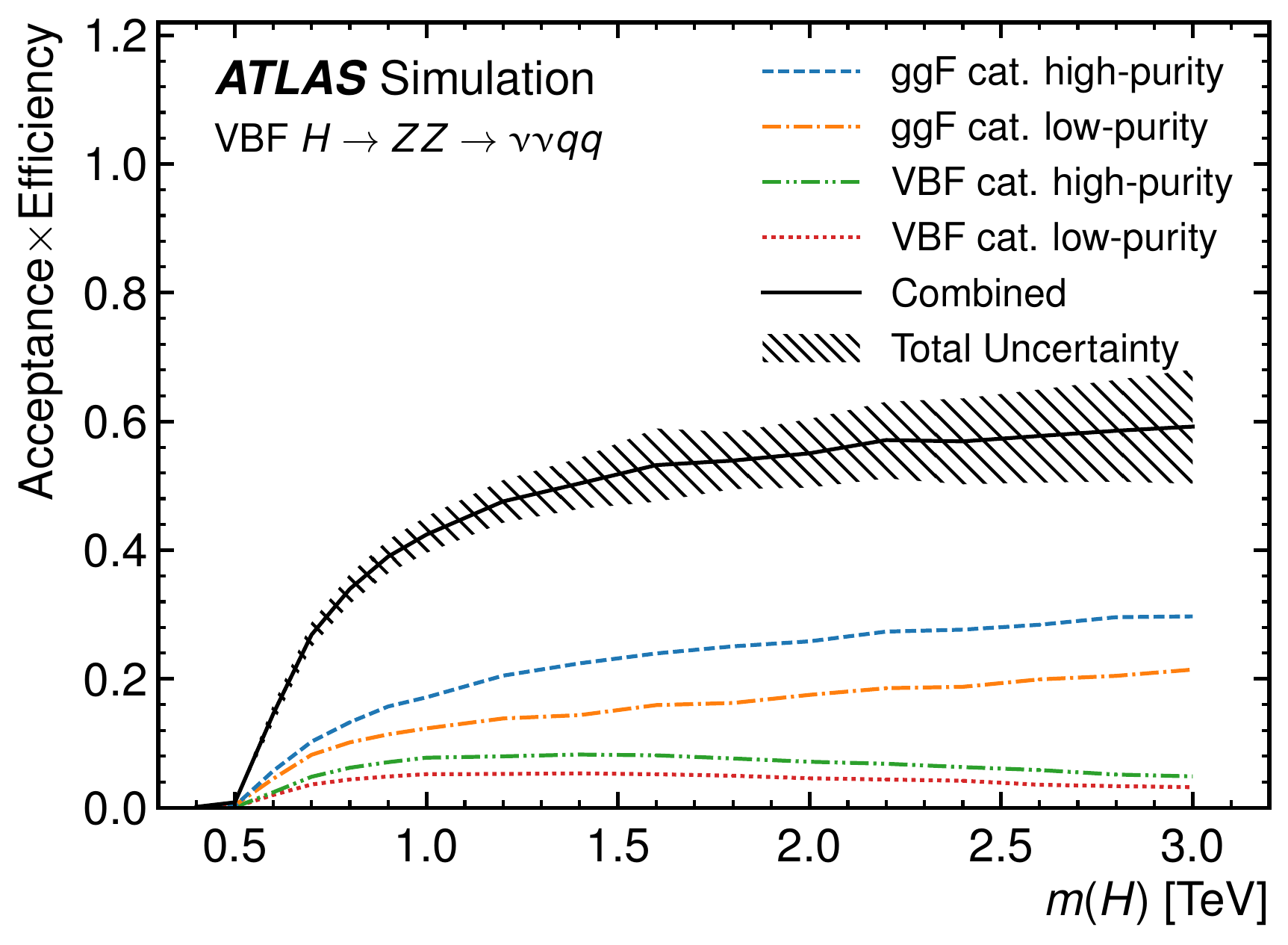}}
  \caption{\label{fig:vvqq_signal} Selection acceptance times efficiency for the $H\to ZZ\to\vvqq$ events from MC simulations as a function of the Higgs boson mass for (a) ggF and (b) VBF production, combining the HP and LP signal regions. The hatched band represents the total statistical and systematic uncertainties.}
\end{figure}

The signal selection efficiency strongly depends on the signal model and the mass of the resonance.
As an example, Figure~\ref{fig:vvqq_signal} shows the selection acceptance times efficiency as a function of $m(H)$ for $H \rightarrow ZV\to\nu\nu qq$. Similar results are obtained for the $W' \to ZW$ and $\Graviton \to ZZ$ searches.

\subsection{Background estimation}
\label{sec:vvqq_background}

The main backgrounds for this search arise from $Z$+jets, $W$+jets and $t\bar{t}$ production. Data control regions are defined to check the modelling of each contribution. As in the \llqq\ search, the shapes of kinematic variables, including the final discriminant $m_{{\mathrm T}}$, are taken from MC simulations.

The normalisation of the $Z$+jets background is determined using the $Z$+jets control region from the \llqq\ search (Section~\ref{sec:llqq_background}). Control regions specific to the $\nu\nu qq$ search are also defined to constrain the $W$+jets and top quark backgrounds. Events in these control regions are selected in the same way as those in the signal regions except that they must have exactly one muon passing the `tight' quality requirement. These events are then split into the top quark and $W$+jets control regions  according to the number of $b$-tagged jets that do not overlap with the large-$R$ jet. Events in the top quark control region must have at least one $b$-tagged jet and events in the $W$+jets control region must not have any $b$-tagged jets.
For the $W$+jets control region, the jet mass requirement used in the signal region selection is inverted, to ensure that there is no resonant diboson signal contamination via $X\to WV \to \ell\nu qq$ decay, while for the top quark control region, the leading large-$R$ jet is required to be consistent with the mass of the $W$ boson.

For the $W$+jets and top quark CRs, the muon is treated as a missing particle to make the boson $\pT$ acceptance more similar to the signal region, and because the trigger does not include muons in the \MET\ calculation. Specifically,
\METnomu (\MPTnomu) is computed by removing the muon \pt\ contribution from the \MET~(\MPT) calculation.
The requirements on \MET\ and \MPT\ in the signal regions are used for \METnomu and \MPTnomu respectively for the control regions.

Table~\ref{tab:vvqq_yields} summarises the numbers of events observed in the data and estimated from background processes in the signal regions from the background-only fit of both the signal and control regions discussed in Section~\ref{sec:results}.
The expected numbers of events from both ggF and VBF production of a 1.6~TeV heavy Higgs boson, assuming $\sigma\times\mathcal{B}(H\to ZZ) = 6$~fb, are also included for comparison.

\begin{table}[!hbt]
  \caption{\label{tab:vvqq_yields} Numbers of events predicted from background processes and observed in the data in the signal regions of the $H\to ZZ\to \vvqq$ search from the background-only fit of both the signal and control regions~(Section~\ref{sec:results}). The numbers of signal events expected from a Higgs boson with mass of 1.6~TeV are also shown. The signal yields are calculated assuming $\sigma\times\mathcal{B}(H\to ZZ) = 6$~fb for both ggF and VBF at 1.6~TeV. The uncertainties combine statistical and systematic contributions.
}
\centering
  \vspace*{0.2cm}
  \footnotesize
  \begin{tabular}{l r@{\,\( \pm \)\,}l r@{\,\( \pm \)\,}l r@{\,\( \pm \)\,}l r@{\,\( \pm \)\,}l r@{\,\( \pm \)\,}l r@{\,\( \pm \)\,}l r@{\,\( \pm \)\,}l r}
  \toprule
    Signal &\multicolumn{4}{c}{$H$ (1.6 TeV)}   & \multicolumn{10}{c}{Background estimates}  & \multirow{2}{*}{Data} \\  \cmidrule(lr){2-5} \cmidrule(lr){6-15}
  regions  & \multicolumn{2}{c}{ggF}  & \multicolumn{2}{c}{VBF}  & \multicolumn{2}{c}{$Z+$jets} &  \multicolumn{2}{c}{$W$+jets}  & \multicolumn{2}{c}{Diboson} & \multicolumn{2}{c}{Top quarks} & \multicolumn{2}{c}{Total} &    \\ \hline \\
    \multicolumn{16}{c}{VBF category} \\ \hline
    HP & $0.49$&$0.15$  & $4.9$&$1.1$  &  $35.0$&$3.3$  &  $25$&$4$      & $6.7$&$1.1$  & $37$&$4$      & $104$&$6$     & 105     \\
    LP & $0.36$&$0.09$  & $3.2$&$0.5$  &  $137$&$10$    &  $94$&$13$     & $12.7$&$2.0$ & $70$&$8$      & $315$&$13$    & 335    \\ \hline \\

    \multicolumn{16}{c}{ggF category} \\ \hline
    HP & $15.5$&$3.3$  & $14.5$&$3.2$  &  $2880$&$80$   &  $1990$&$100$  & $490$&$60$   & $1500$&$90$   & $6870$&$70$   & 6888   \\
    LP & $10.6$&$1.5$  & $9.7$&$1.4$   &  $11160$&$210$ &  $7200$&$300$  & $830$&$90$   & $2710$&$180$  & $21910$&$140$ & 21936  \\
  \bottomrule
  \end{tabular}
\end{table}

%% file: tex/systematics.tex
Systematic uncertainty sources impacting the search can be divided into three categories:
experimental uncertainties related to the detector or to the reconstruction algorithms,
uncertainties in the estimations of background contributions, and uncertainties in modelling the signal. Unless explicitly stated, the uncertainties quoted below are the uncertainties in the quantities themselves, not the impact on the search sensitivity.

The uncertainty in the integrated luminosity of the dataset is determined to be 3.2\%.
It is derived, following a methodology similar to the one detailed in Ref.~\cite{Aaboud:2016hhf},
from a preliminary calibration of the luminosity scale using a pair of $x$--$y$ beam-separation scans performed in August 2015 and May 2016.
This uncertainty is applied to the normalisation of the signal and also to background contributions whose normalisations are derived from MC simulations.
A variation in the pile-up reweighting of MC events is included to cover the uncertainty in the ratio of the predicted and measured inelastic cross section in Ref.~\cite{Aaboud:2016mmw}.

The efficiencies of the lepton triggers for events with selected leptons are high, nearly 100\% in the electron channel and approximately 96\% in the muon channel.  The corresponding uncertainties are negligible.
For the selection used in the \vvqq\ search, the efficiency of the $\met$ trigger is also close to 100\% with negligible associated uncertainty.
The modelling of the electron and muon reconstruction, identification and isolation efficiencies
is studied with a tag-and-probe method using $Z\to\ell\ell$ events in data and simulation at \cms\
\cite{Muon_perf_2015,Elec_ID_2015}.
Small corrections are applied to the simulation to better model the performance seen in data.
These corrections have associated uncertainties of the order of 1\%.
For the $\vvqq$ search, the uncertainty is instead in the efficiency of the lepton veto, and is at the sub-percent level.
Uncertainties in the lepton energy (or momentum) scale and resolution are also taken into account.

Uncertainties in the jet energy scale and resolution for small-radius jets
are estimated using MC simulation and {\em in situ} techniques~\cite{Aaboud:2017jcu}.
For central jets ($|\eta|<2.0$), the total uncertainty in the jet energy scale ranges from about 6\%
for jets with $\pt = 25~\gev$ to about 2\% for $\pt = 1~\tev$.
There is also an uncertainty in the jet energy resolution~\cite{Run2JER},
which ranges from 10\% to 20\% for jets with a $\pt$ of $20~\gev$ to less than 5\% for jets with $\pt>200~\gev$.
Uncertainties in the lepton and jet energy scales and resolutions are propagated into
the uncertainty in $\met$.  
Uncertainties in the energy scale and resolution of the track soft term
are also propagated into the uncertainty in $\met$~\cite{ATL-PHYS-PUB-2015-023}, which is about 2\% for the \vvqq\ search.
Uncertainties in the efficiency for tagging $b$-jets and in the rejection factor
for light jets are determined from $\ttbar$ samples~\cite{ATL-btag-PHYS-PUB-2015-039,ATL-PHYS-PUB-2015-022}. 

The uncertainties in the scale of the large-$R$ jet $\pT$, mass and \DTwoBetaOne\
are of the order of 2--5\%. 
They are estimated using comparisons of data and simulation in Ref.~\cite{Aad:2015rpa}.
An absolute uncertainty of 2\% is assigned to the large-$R$ jet energy resolution, and relative uncertainties of 20\% and 15\% are assigned to the resolution of the large-$R$ jet mass and \DTwoBetaOne, respectively.

For the \llqq\ search the uncertainty in the spectrum of the final discriminant, \mllj\ or \mlljj, for $Z$+jets,
is assessed by comparing the shape difference between data and MC simulations in the control regions. The uncerainty is found to be approximately 20\% for \mlljj\ and varies from 3\% at low mass to 20\% at high mass for \mllj. 
For the $\vvqq$ search the uncertainty in the modelling of the $m_{\mathrm T}$ distribution for $Z$+jets and $W$+jets production is assessed by comparing the nominal \textsc{Sherpa} MC sample with alternative samples generated by doubling or halving the renormalisation, resummation and factorisation scales independently and varying the matrix-element matching. The uncertainties are of the order of 5--30\%.
The approach is different from that used in the \llqq\ search, since it is difficult to obtain a pure control region for $Z$+jets or $W$+jets in the \vvqq\ search.

An uncertainty in the shape of the  \mllj\ or $m_{\mathrm T}$
distribution for the $t\bar{t}$  background is derived by comparing the {\textsc Powheg-Box} sample with the distribution obtained using {\textsc MadGraph5\_aMC@NLO 2.2.2}. This uncertainty is found to be approximately 15\%. 
Additional systematic uncertainties are estimated by comparing the
nominal sample showered with {\textsc Pythia} 6.428 using the P2012 tune to
one showered with {\textsc Herwig++} 2.7.1~\cite{Herwig} and using the UEEE5 underlying-event tune. The uncertainty varies between 10--40\%.
Samples of $t \bar{t}$ events with the factorisation and renormalisation
scales doubled or halved are compared to the nominal, and the differences observed are taken as an additional uncertainty. These uncertainties are small, typically less than 3\%.

The uncertainties in the diboson cross sections are estimated to be 10\%~\cite{Gleisberg:2008ta,Diboson_NLO},
and the shape uncertainty is obtained by comparing MC samples generated by \textsc{Sherpa} and \textsc{Powheg-Box}.

The signals acceptance uncertainty due to PDF uncertainties is estimated by taking the uncertainty
from the PDF error sets, CT10 for the heavy Higgs boson signal and NNPDF23LO for the HVT and bulk RS signals, 
and adding it in quadrature to the acceptance difference obtained with the use of
alternative PDF sets: MMHT2014LO~\cite{MMH2014} and NNPDF30NLO~\cite{Ball:2014uwa} for the Higgs boson, and
\textsc{CT10} and MMHT2014LO for the HVT and bulk RS graviton. For the HVT signal, the uncertainty ranges from 1\%~(2\%) to 6\%~(12\%) for the \llqq\ (\vvqq) search depending on the mass being tested. Similar results are obtained for the heavy Higgs boson signal, while for the bulk RS signal, the uncertainty is generally less than 1\%.
The signal acceptance uncertainty due to initial- and final-state radiation (ISR/FSR) is estimated by varying relevant parameters in the A14-NNPDF tune~\cite{A14_tune} for the HVT and bulk RS signals. This uncertainty ranges from less than 1\% to 4\%~(6\%) for the \llqq\ (\vvqq) search depending on the mass being tested.
The effect of the QCD scale uncertainty on the heavy Higgs boson signal acceptance is estimated by varying the factorisation and renormalisation scales. It ranges from 1\%~(2\%) to 4\%~(8\%) for the \llqq\ (\vvqq) search depending on the mass being tested.

%% file: tex/results.tex
\subsection{Statistical and fit procedures}

The statistical analysis is based on the framework described in Refs.~\cite{Aad:2012an,Moneta:2010pm,Verkerke:2003ir}.
A binned likelihood function $\mathcal{L}(\mu,\theta)$ is constructed as a product of Poisson probability terms over all bins of the fit templates considered in the search. This function depends on the signal-strength parameter $\mu$, a multiplicative factor applied to the theoretical signal production cross section, and $\theta$, a set of nuisance parameters that encode the effects of systematic uncertainties in the signal and expected backgrounds, described in Section~\ref{sec:systematics}. The binning is chosen based on the expected mass resolution and numbers of events. The nuisance parameters are either free to float, or constrained using Gaussian terms defined by external studies.
The likelihood function for the combination of the \llqq\ and \vvqq\ channels is the product of the Poisson likelihoods of these individual channels. However, only one constraint term per common nuisance parameter is included in the product.

For each signal process considered, a simultaneous maximum-likelihood fit is performed to the observed distributions of the final discriminants in the signal regions, $m_{\ell\ell J}$ or $m_{\ell\ell jj}$ for the $X\to ZV\to\ell\ell qq$ search and $m_{\mathrm T}$ for the $X\to ZV\to\nu\nu qq$ search,  to extract the signal rate information.
The shapes of the signal distributions are interpolated using the moment-morphing method~\cite{Baak:2014fta} for intermediate mass points where simulated signal samples are not available. The product of acceptance and efficiency for each interpolated signal mass point is obtained from an interpolation with cubic splines between the simulated mass points.
The $Z$+jets, $W$+jets and top quark control regions are included in
the fit's likelihood calculation
with one bin per region, i.e., using only their event-count information. 
Background contributions, including their shapes in the signal regions, are taken from MC simulations. However, they are allowed to vary independently within their uncertainties in each bin. Moreover, normalisation scale factors~(SFs) are applied to  the MC estimates of the $Z$+jets, $W$+jets and top quark contributions. These scale factors are free parameters in the fit and are therefore constrained by the data in both the signal and control regions. The diboson contribution is constrained to the theoretical estimate within the corresponding uncertainties.

In general, one SF is introduced per control region for its intended background component with the following exceptions to take into account different MC modellings in the different phase spaces of the same background component. One common $Z$+jets SF is used for HP and LP regions of the merged selection of both the $ZV\to\llqq$ and $ZV\to\vvqq$ searches. However, independent SFs are used for the VBF and ggF categories. For the resolved $ZV\to\llqq$ selection, two independent $Z$+jets SFs are used for the $b$-tagged and untagged regions: one for the $Z$+heavy-flavour component and the other for the $Z$+light-jet component (Section~\ref{sec:samples}). The $ZV\to\llqq$ search defines a top quark control region and consequently applies one top quark SF to all regions. The $ZV\to\vvqq$ search shares the $Z$+jets control regions with $ZV\to\llqq$ and hence applies the same SFs to its $Z$+jets background contributions. 
The $ZV\to\vvqq$ search has its own $W$+jets and top quark control regions. Similar to $Z$+jets, one common $W$+jets SF is used for HP and LP regions, but independent SFs are used for the VBF and ggF categories. All the top quark control regions in the $ZV\to\vvqq$ search share one common SF and it is distinguished from the top quark SF in the $ZV\to\llqq$ search.

The test statistic $q_{\mu}$ is defined as the profile likelihood ratio~\cite{ATL-PHYS-PUB-2011-011},
$q_{\mu}=-2\mathrm{ln}(\mathcal{L}(\mu,\hat{\hat{\theta}}_{\mu})/ \mathcal{L}(\hat{\mu}, \hat{\theta}))$,
where $\hat{\mu}$ and $\hat{\theta}$ are the values of the parameters that maximise the likelihood function (with the constraint 0$\leq  \hat{\mu} \leq \mu$), and $\hat{\hat{\theta}}_{\mu}$ are the values of the nuisance parameters that maximise the likelihood function for a given value of $\mu$. The test statistic $q_{\mu}$ is used to measure the compatibility of the observed data with the background-only hypothesis.
In the absence of a signal, constraints on the production of a heavy resonance decaying into $ZV$ pairs are derived.
The exclusion limits are calculated with a modified frequentist method \cite{CLs_2002}, also known as CL$_{\mathrm{s}}$,
in the asymptotic approximation~\cite{Cowan:2010js} for resonance masses below 2~TeV. Above 2~TeV, the small number of events makes the asymptotic approximation unreliable and the limits are calculated using pseudo-experiments. All limits are set at the 95\% confidence level (CL).

\subsection{Limits on the production of heavy resonances}

\begin{figure}[!h!btp]
\centering
  \subfigure[]{\includegraphics[width=0.495\textwidth]{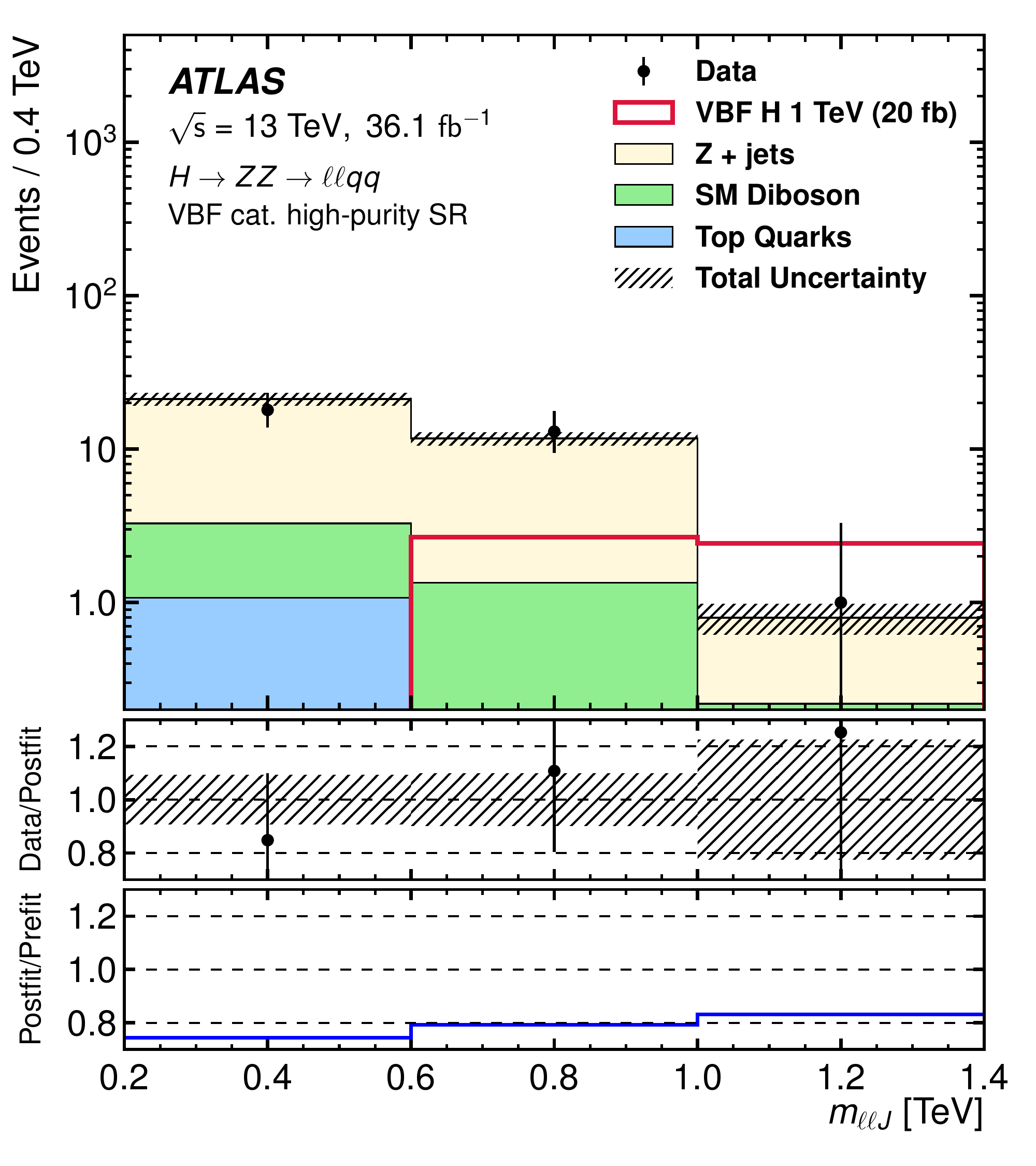}}
  \subfigure[]{\includegraphics[width=0.495\textwidth]{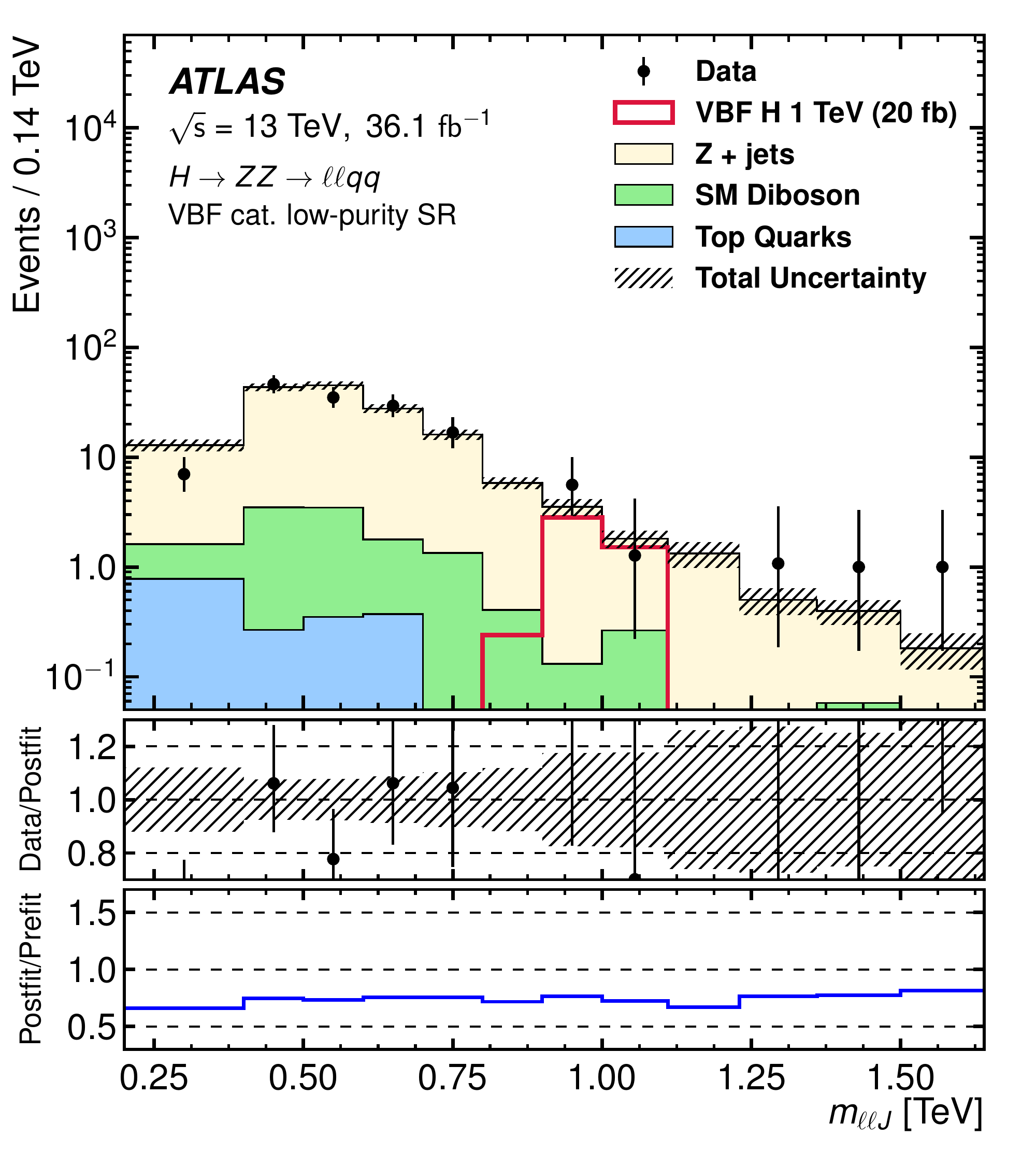}} \\
  \subfigure[]{\includegraphics[width=0.495\textwidth]{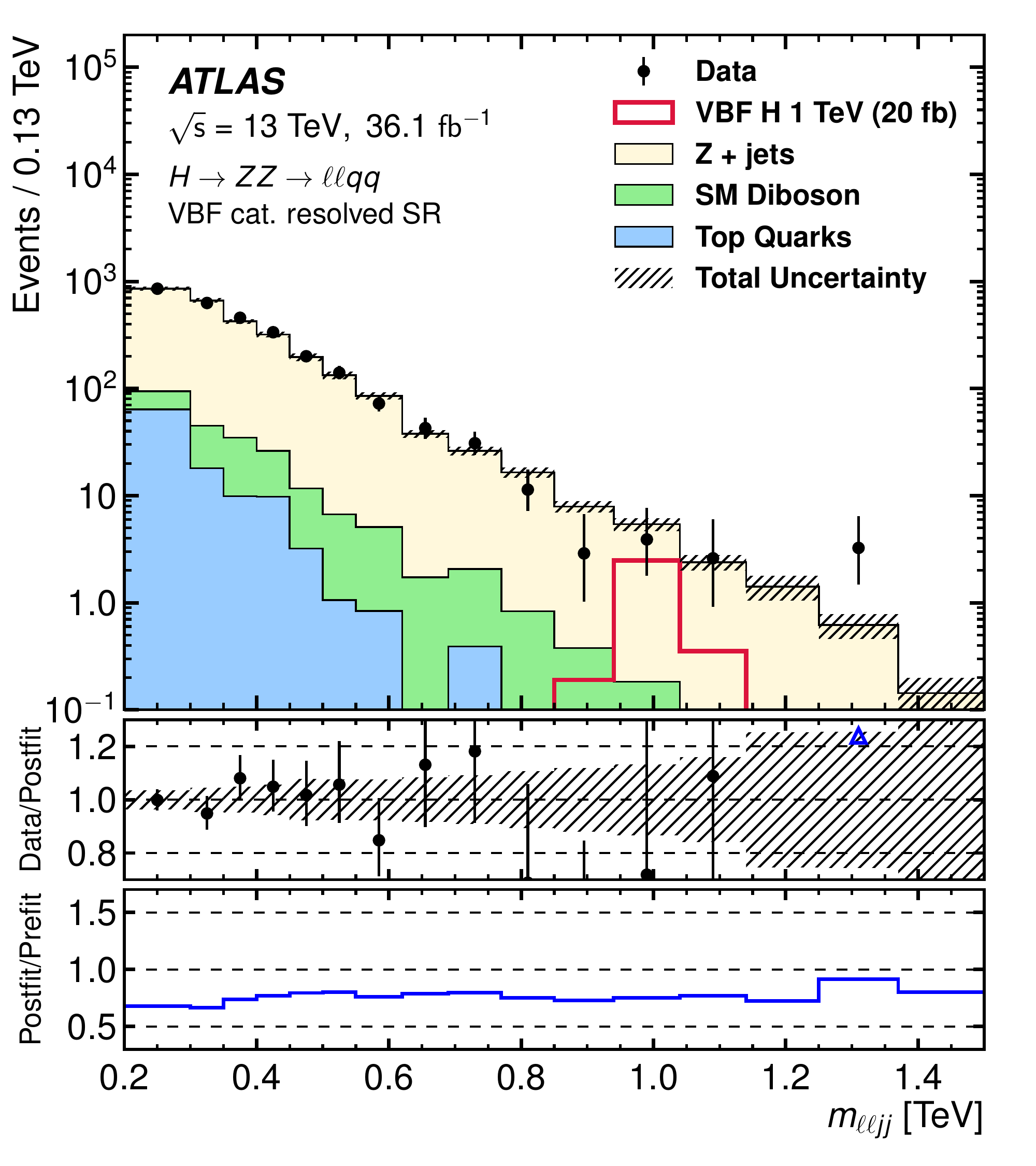}}
  \caption{\label{fig:llqq_mlljjJ_VBF} Comparisons of the observed data and expected background distributions of the final discriminants of the VBF category for the  $H\to ZZ\to\ell\ell qq$ search:  $\mllJ$ of (a) high-purity and (b) low-purity signal regions; (c) $\mlljj$ of the resolved signal region.
For illustration, expected distributions from the VBF production of a 1~TeV Higgs boson with $\sigma\times\mathcal{B}(H\to ZZ)=20$~fb are also shown.
The middle panes show the ratios of the observed data to the background predictions. The uncertainty in the total background prediction, shown as bands, combines statistical and systematic contributions. The blue triangles in the middle panes  indicate bins where the ratio is nonzero and outside the vertical range of the plot. The bottom panes show the ratios of the post-fit and pre-fit background predictions.}
\end{figure}

\begin{figure}[!h!btp]
\centering
  \subfigure[]{\includegraphics[width=0.495\textwidth]{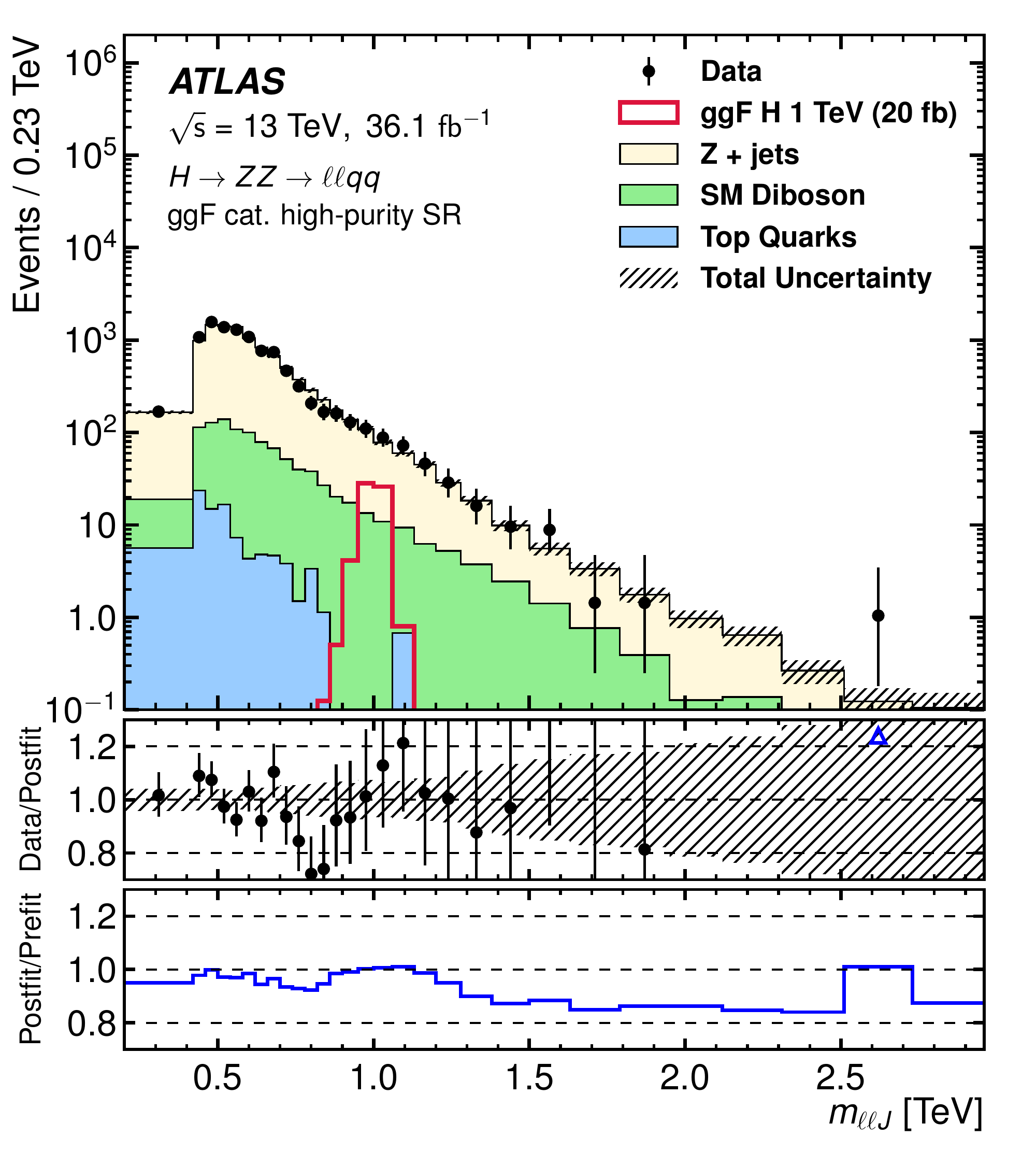}}
  \subfigure[]{\includegraphics[width=0.495\textwidth]{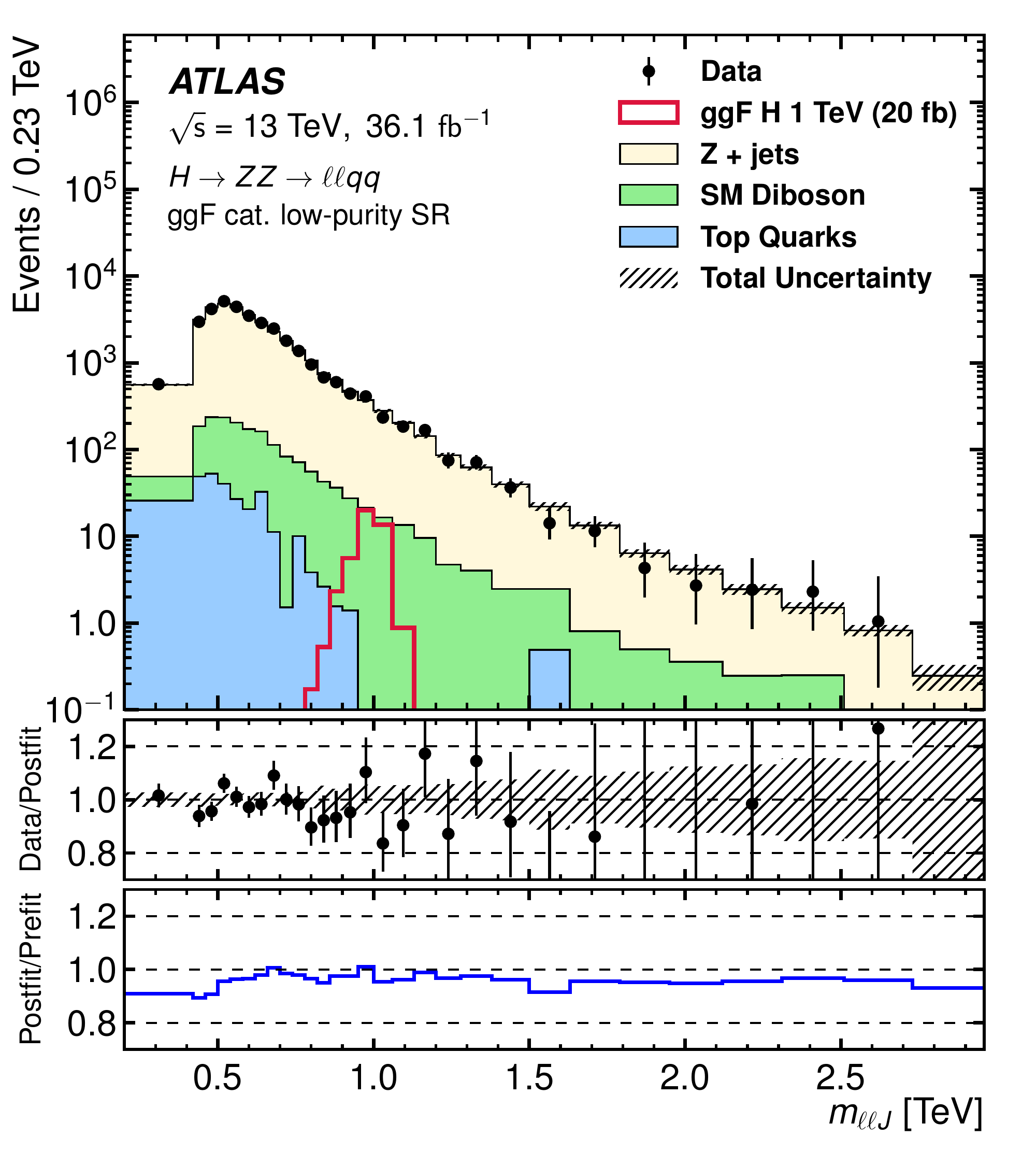}}
  \subfigure[]{\includegraphics[width=0.495\textwidth]{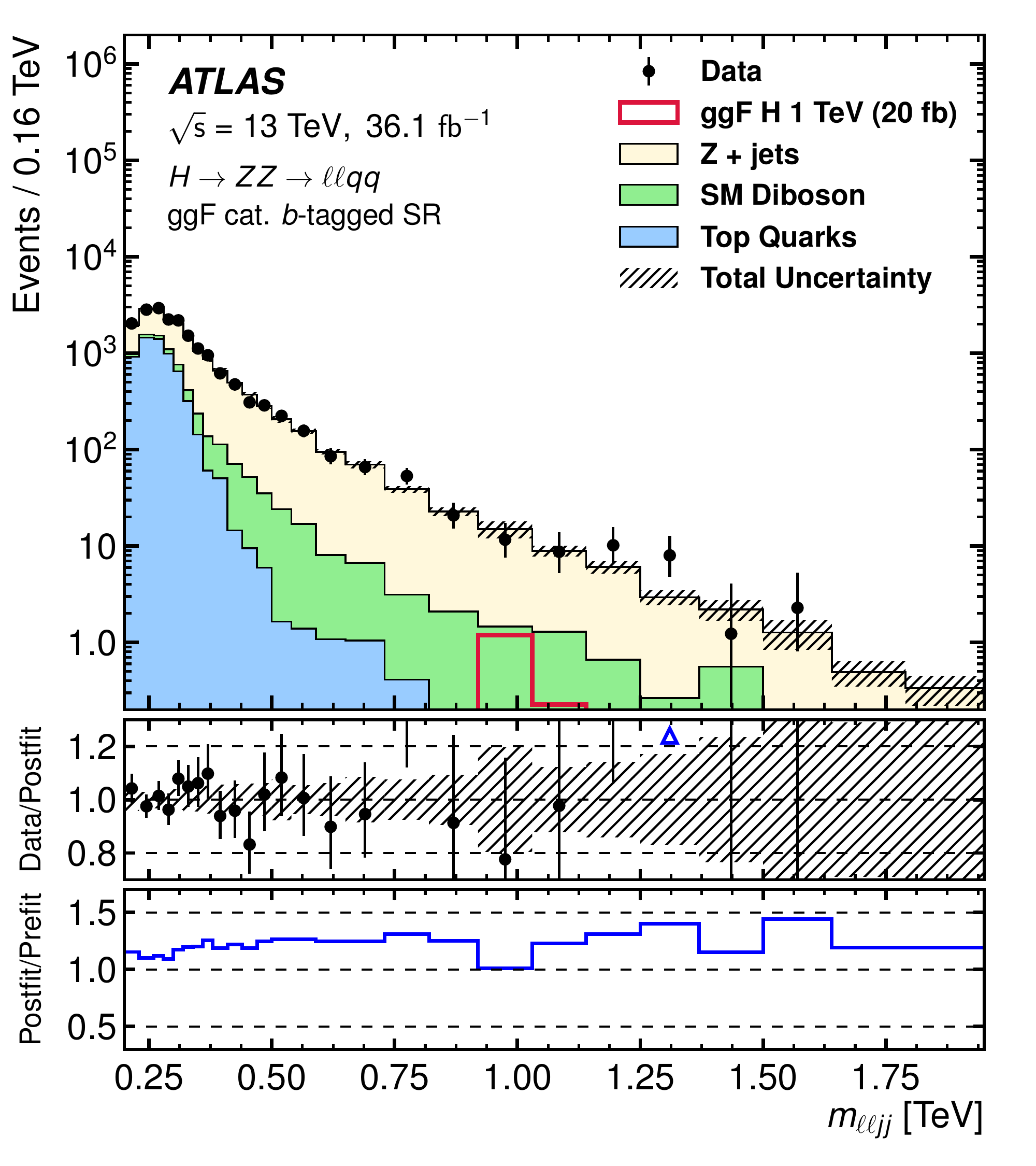}}
  \subfigure[]{\includegraphics[width=0.495\textwidth]{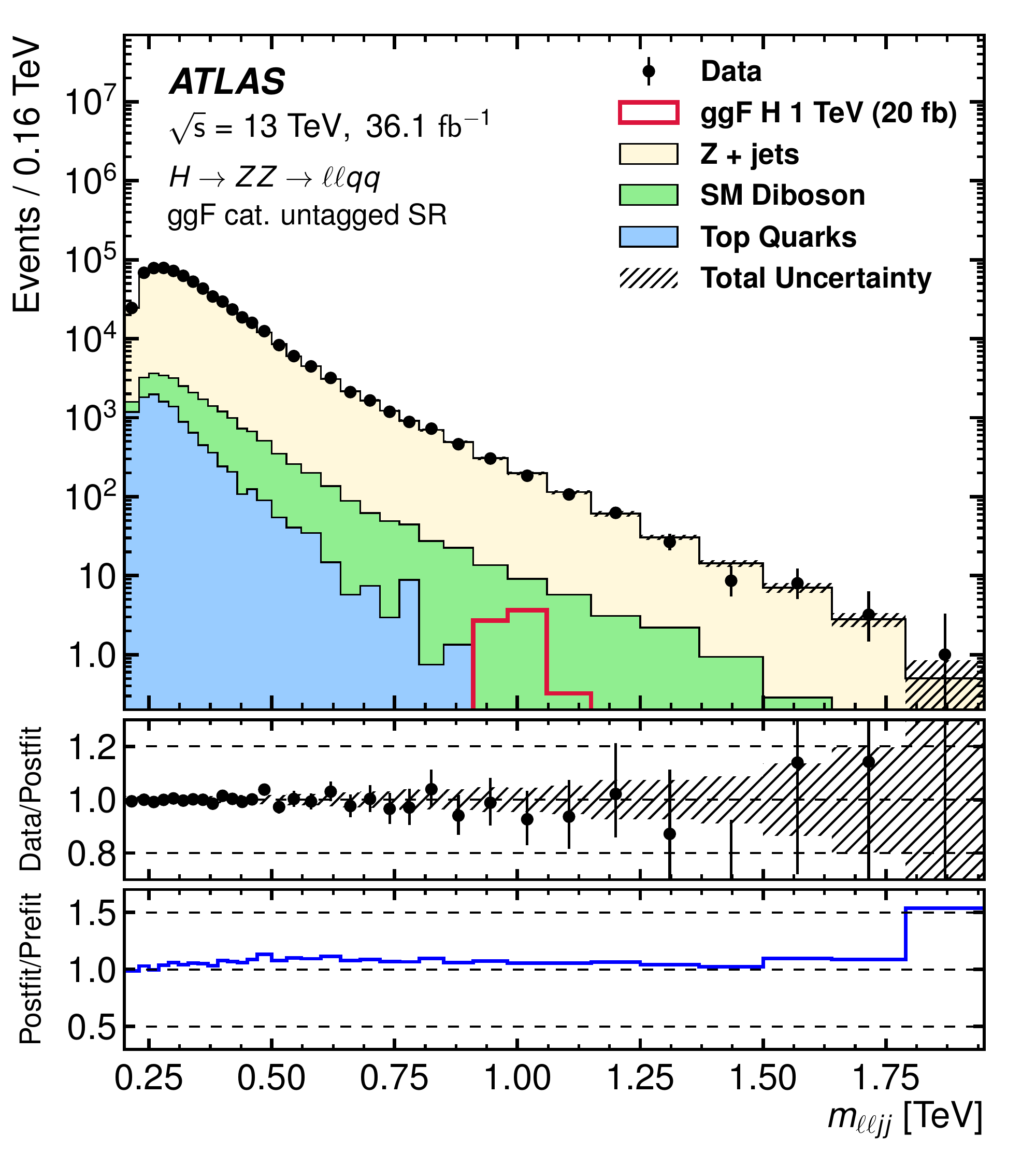}} \\
  \caption{\label{fig:llqq_mlljjJ_ggF} Comparisons of the observed data and expected background distributions of the final discriminants  of the ggF category for the  $H\to ZZ\to\ell\ell qq$ search: $m_{\ell\ell J}$ of (a) high-purity and (b) low-purity signal regions; $m_{\ell\ell jj}$  of (c) $b$-tagged and (d) untagged signal regions.
For illustration, expected distributions from the ggF production of a 1~TeV Higgs boson with $\sigma\times\mathcal{B}(H\to ZZ)=20$~fb are also shown. The middle panes show the ratios of the observed data to the background predictions.
The uncertainty in the total background prediction, shown as bands, combines statistical and systematic contributions. The blue triangles in the middle panes  indicate bins where the ratio is nonzero and outside the vertical range of the plot. The bottom panes show the ratios of the post-fit and pre-fit background predictions.}
\end{figure}

\begin{figure}[!h!btp]
\centering
  \subfigure[]{\includegraphics[width=0.495\textwidth]{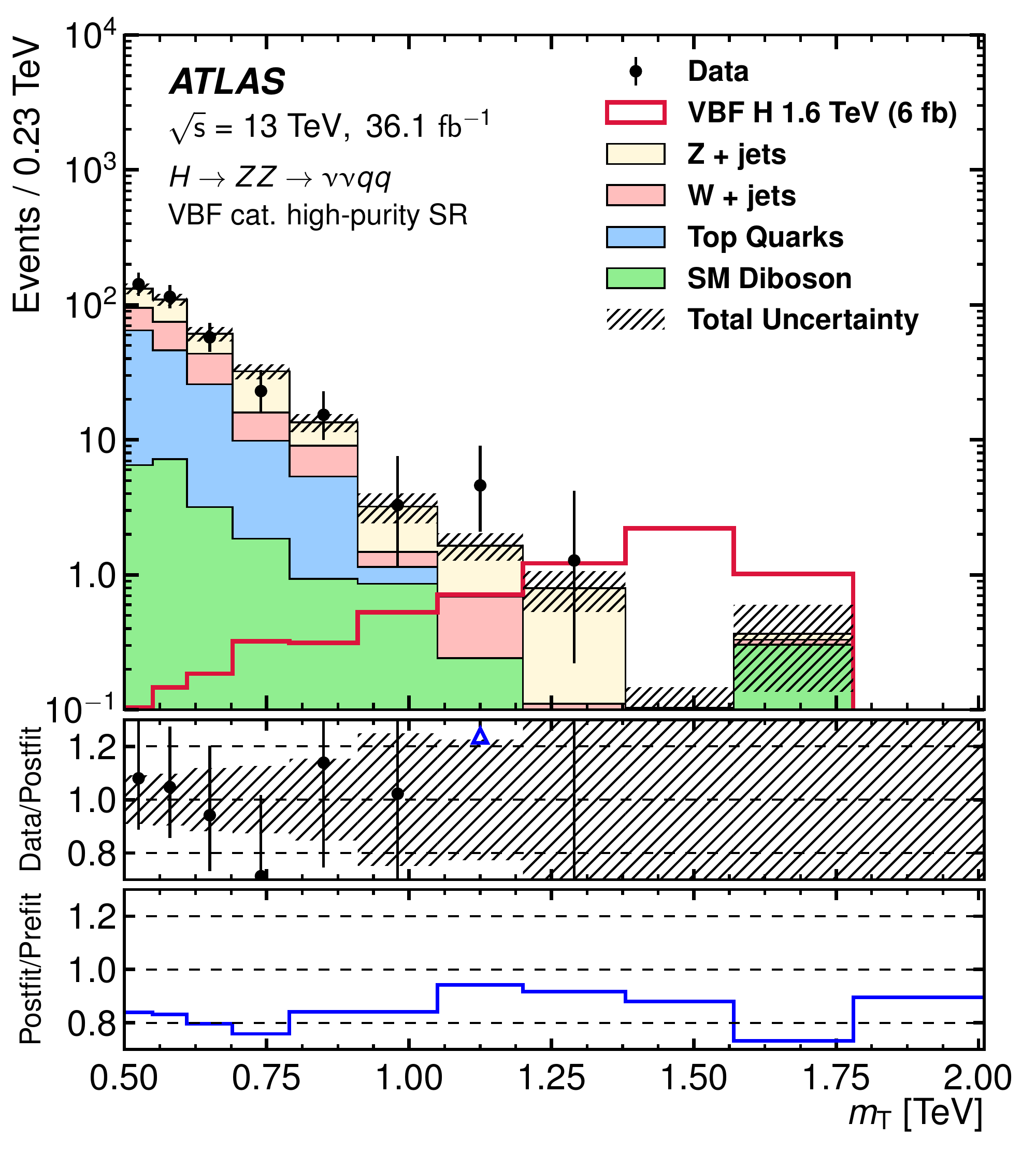}}
  \subfigure[]{\includegraphics[width=0.495\textwidth]{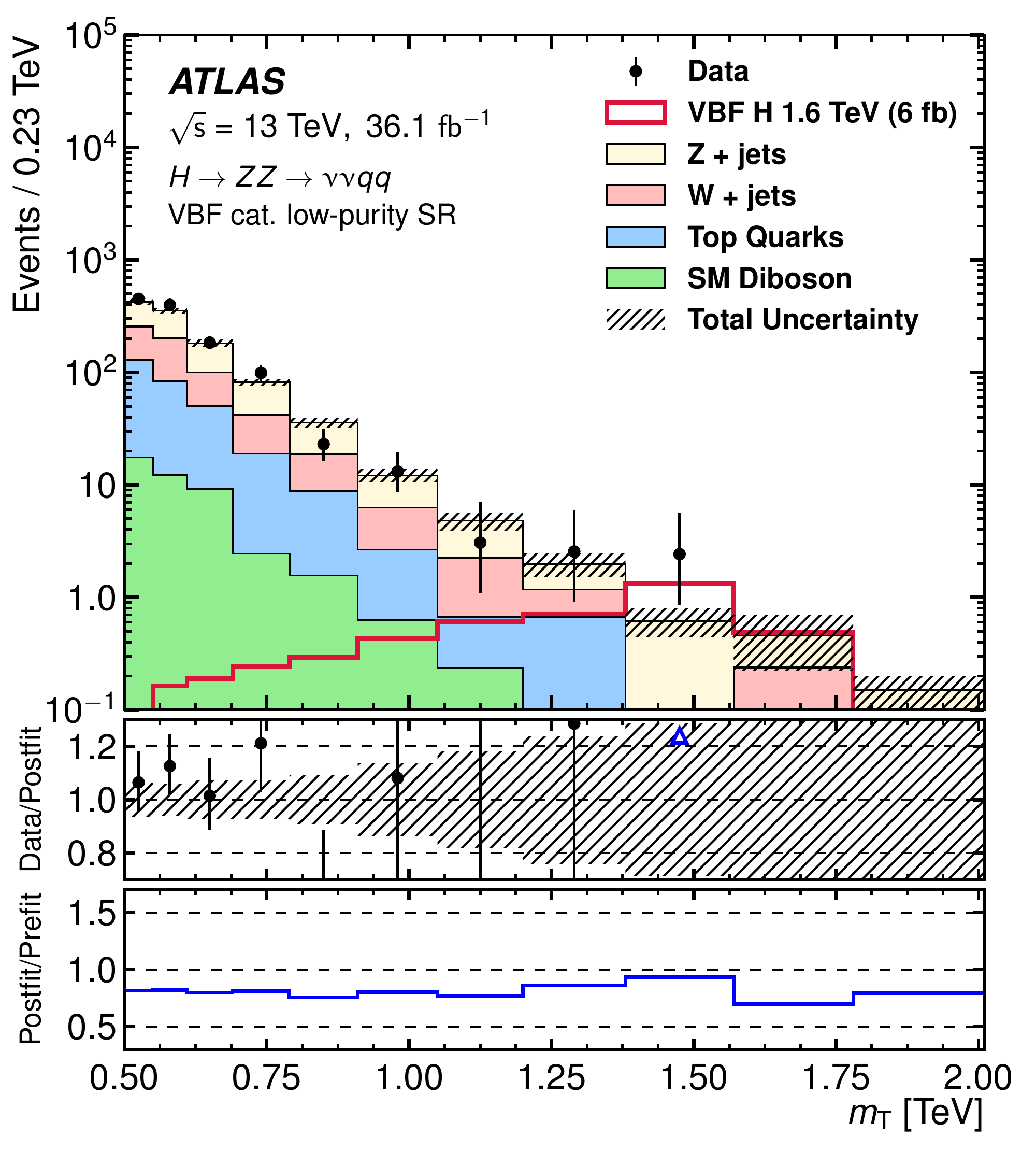}} 

  \caption{\label{fig:vvqq_SR_VBF} Comparisons of the observed data and expected background distributions of $m_{{\mathrm T}}$ in the VBF category of the  $H\to ZZ\to\vvqq$ search:  (a) high-purity and (b) low-purity signal regions.
For illustration, expected distributions from the VBF production of a 1.6~TeV Higgs boson with $\sigma\times\mathcal{B}(H\to ZZ)=6$~fb are also shown. The middle panes show the ratios of the observed data to the  background predictions.
The uncertainty in the total background prediction, shown as bands, combines statistical and systematic contributions. The blue triangles in the middle panes  indicate bins where the ratio is nonzero and outside the vertical range of the plot. The bottom panes show the ratios of the post-fit and pre-fit background predictions.}
\end{figure}

\begin{figure}[!h!btp]
\centering
  \subfigure[]{\includegraphics[width=0.495\textwidth]{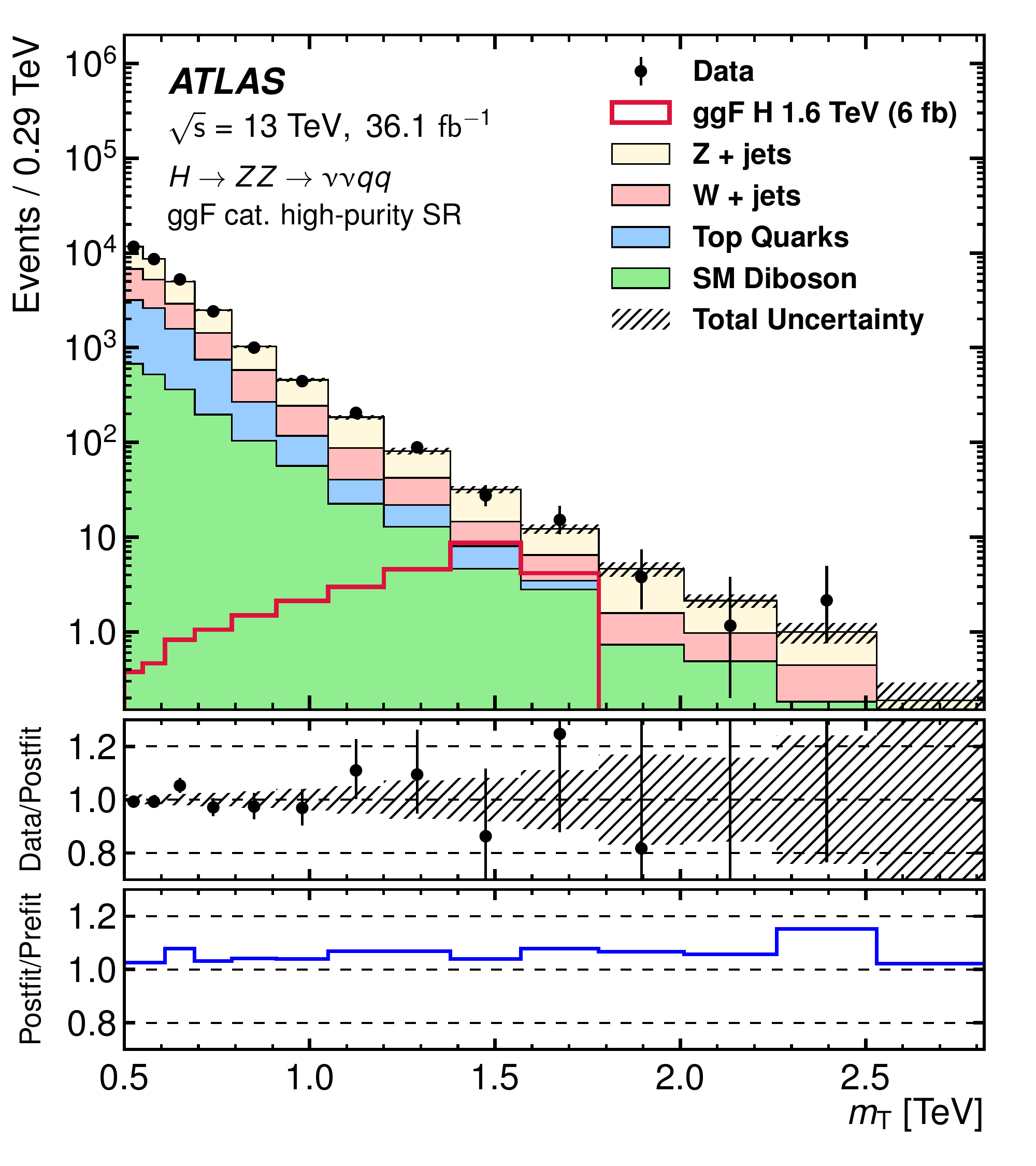}}
  \subfigure[]{\includegraphics[width=0.495\textwidth]{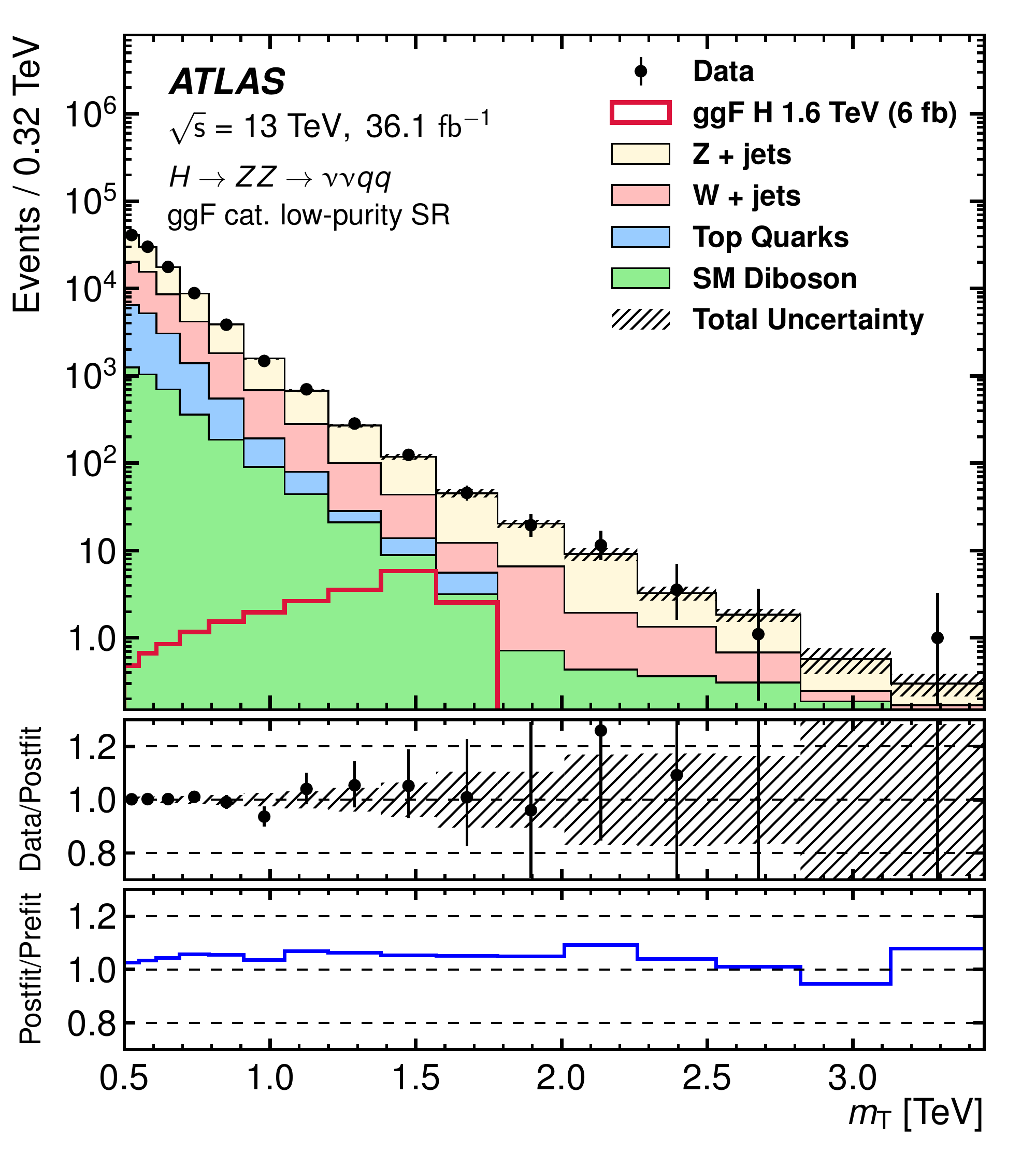}} 

  \caption{\label{fig:vvqq_SR_ggF} Comparisons of the observed data and expected background distributions of $m_{{\mathrm T}}$ in the ggF category of the $H\to ZZ\to\vvqq$ search:  (a) high-purity and (b) low-purity signal regions.
For illustration, expected distributions from the ggF production of a 1.6~TeV Higgs boson with $\sigma\times\mathcal{B}(H\to ZZ)=6$~fb are also shown. The middle panes show the ratios of the observed data to the background predictions.
The uncertainty in the total background prediction, shown as bands, combines statistical and systematic contributions. The blue triangles in the middle panes  indicate bins where the ratio is nonzero and outside the vertical range of the plot. The bottom panes show the ratios of the post-fit and pre-fit background predictions.}
\end{figure}

\begin{figure}[!h!btp]
\centering
   \includegraphics[width=0.65\textwidth]{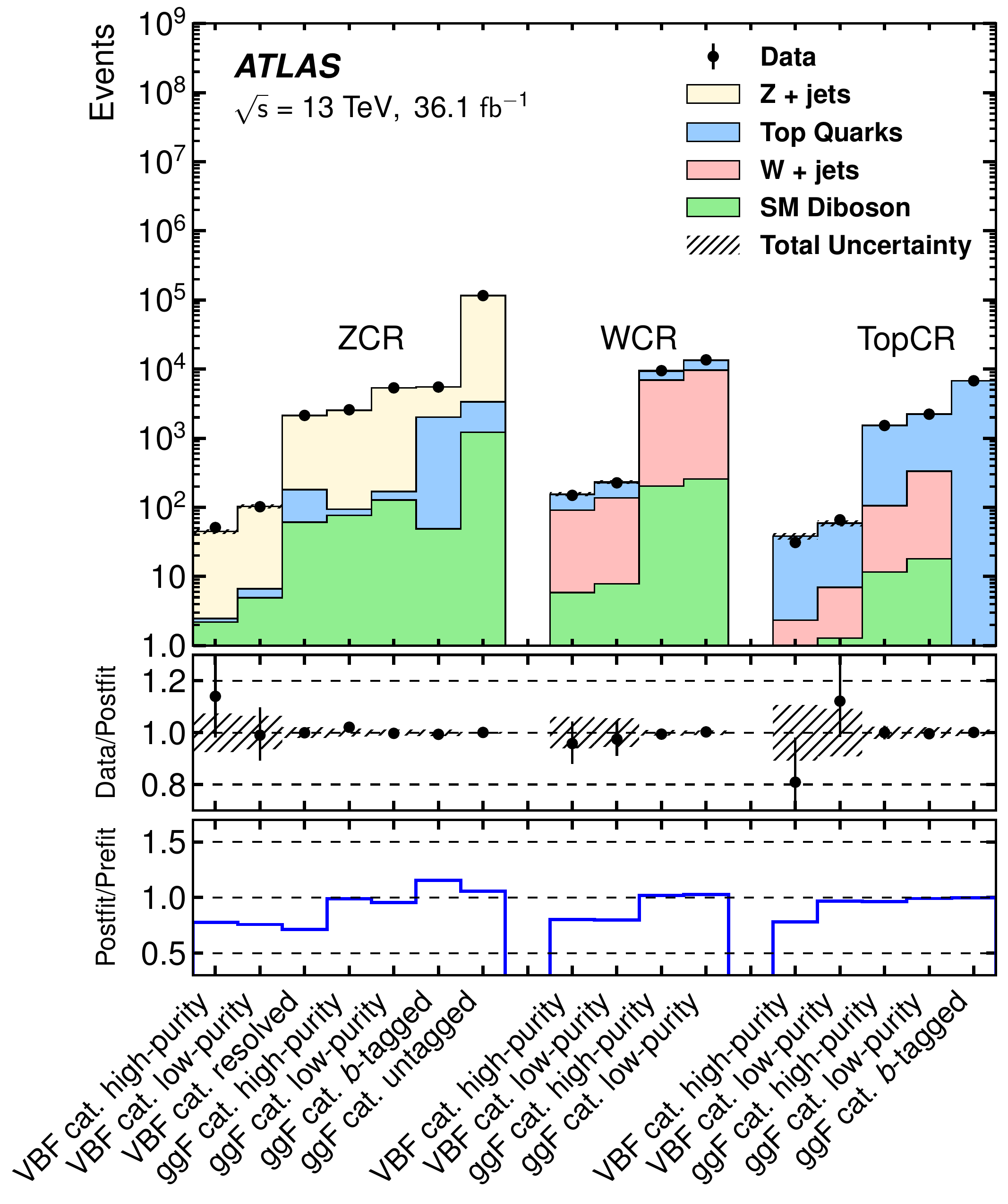}
  \caption{\label{fig:comb_CR}Comparisons of the observed data and expected background event yields in each control region. The middle pane shows the ratios of the observed data to the post-fit background predictions.
The uncertainty in the total background prediction, shown as bands, combines statistical and systematic contributions. The bottom pane shows the ratios of the post-fit and pre-fit background predictions.}
\end{figure}

The observed distributions of $\mllJ$ or $\mlljj$ from the $H\to ZZ\to \llqq$ search are compared with the background estimates in Figure~\ref{fig:llqq_mlljjJ_VBF} for the three signal regions of the VBF category and in Figure~\ref{fig:llqq_mlljjJ_ggF} for the four signal regions of the ggF category. Similarly,  Figures~\ref{fig:vvqq_SR_VBF} and~\ref{fig:vvqq_SR_ggF} show the $m_{\text T}$ distributions from the $H\to ZZ\to \vvqq$ search in the signal regions of the VBF and ggF categories, respectively.
There are no events in the data beyond the ranges of the distributions shown.
Background contributions shown are obtained from background-only fits described previously. The numbers of events observed and estimated in control regions are summarised in Figure~\ref{fig:comb_CR}.

The data distributions are reasonably well reproduced by the estimated background contributions in all these distributions. Similar levels of agreement are observed for the $W'\to ZW$ and $\Graviton\to ZZ$ searches. The largest difference between the observed data and the SM background prediction is in the $H \to ZZ\to \llqq$
search, high-purity signal region, as shown in Figure~\ref{fig:llqq_mlljjJ_ggF}(a). A deficit in data is seen around $\mllJ$ of 800 GeV, with a local significance of $3.0\sigma$, and by taking into account the look-elsewhere effect~\cite{Gross:2010qma} in the full $\mllJ$ spectrum, a global significance of $1.9\sigma$.
Upper limits are set on the product of the production cross section of new resonances and their decay branching ratio to $ZV$, $\sigma\times\mathcal{B}(X\to ZV)$.

\clearpage

\begin{figure}[h!bt]
\centering
  \subfigure[ggF $H\to ZZ$]{\includegraphics[width=0.495\textwidth]{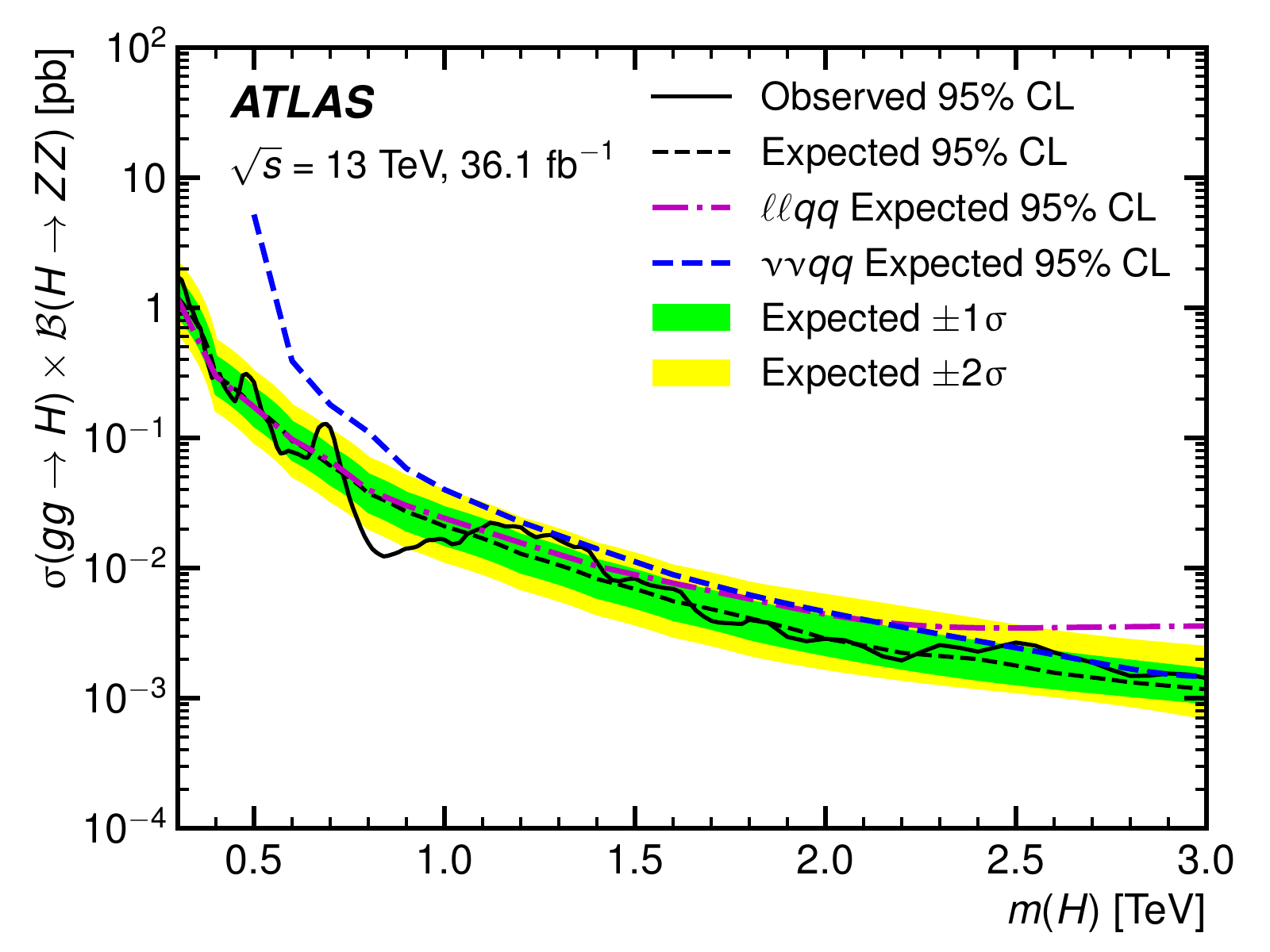}}
  \subfigure[VBF $H\to ZZ$]{\includegraphics[width=0.495\textwidth]{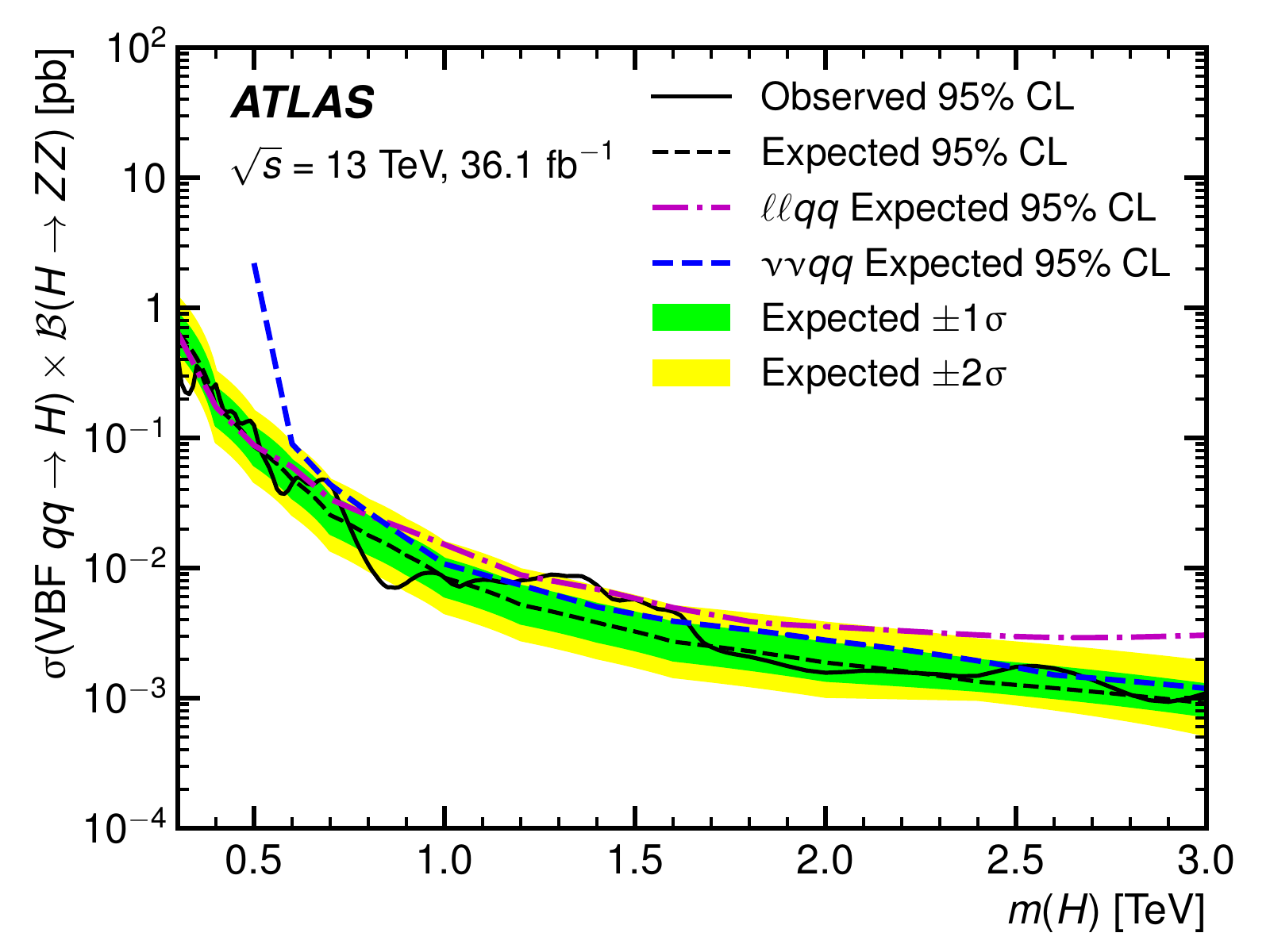}}  \\
  \caption{\label{fig:limits_Higgs}
   Observed (black solid curve) and expected (black dashed curve) 95\% CL upper limits on $\sigma\times\mathcal{B}(H\to ZZ)$ at $\sqrt{s}=13$~TeV for the (a) ggF and (b) VBF production of a heavy Higgs boson as a function of its mass, combining $\llqq$ and $\vvqq$ searches. Limits expected from individual searches (dashed curves in blue and magenta) are also shown for comparison. Limits are calculated in the asymptotic approximation below 2~TeV and are obtained from pseudo-experiments above that.
The green (inner) and yellow (outer) bands represent the $\pm 1\sigma$ and $\pm 2\sigma$ uncertainty in the expected limits. }
\end{figure}

Limits are presented for both the ggF and VBF productions of $H\to ZZ$ in Figure~\ref{fig:limits_Higgs} in the resonance mass range between 300~GeV and 3~TeV. The observed limit on $\sigma\times\mathcal{B}(H\to ZZ)$ varies from 1.7 (0.42) pb at 300~GeV to 1.4 (1.1) fb at 3~TeV for ggF (VBF) $H\to ZZ$ production.

\begin{figure}[h!bt]
\centering
  \subfigure[DY  $W'\to ZW$]{\includegraphics[width=0.495\textwidth]{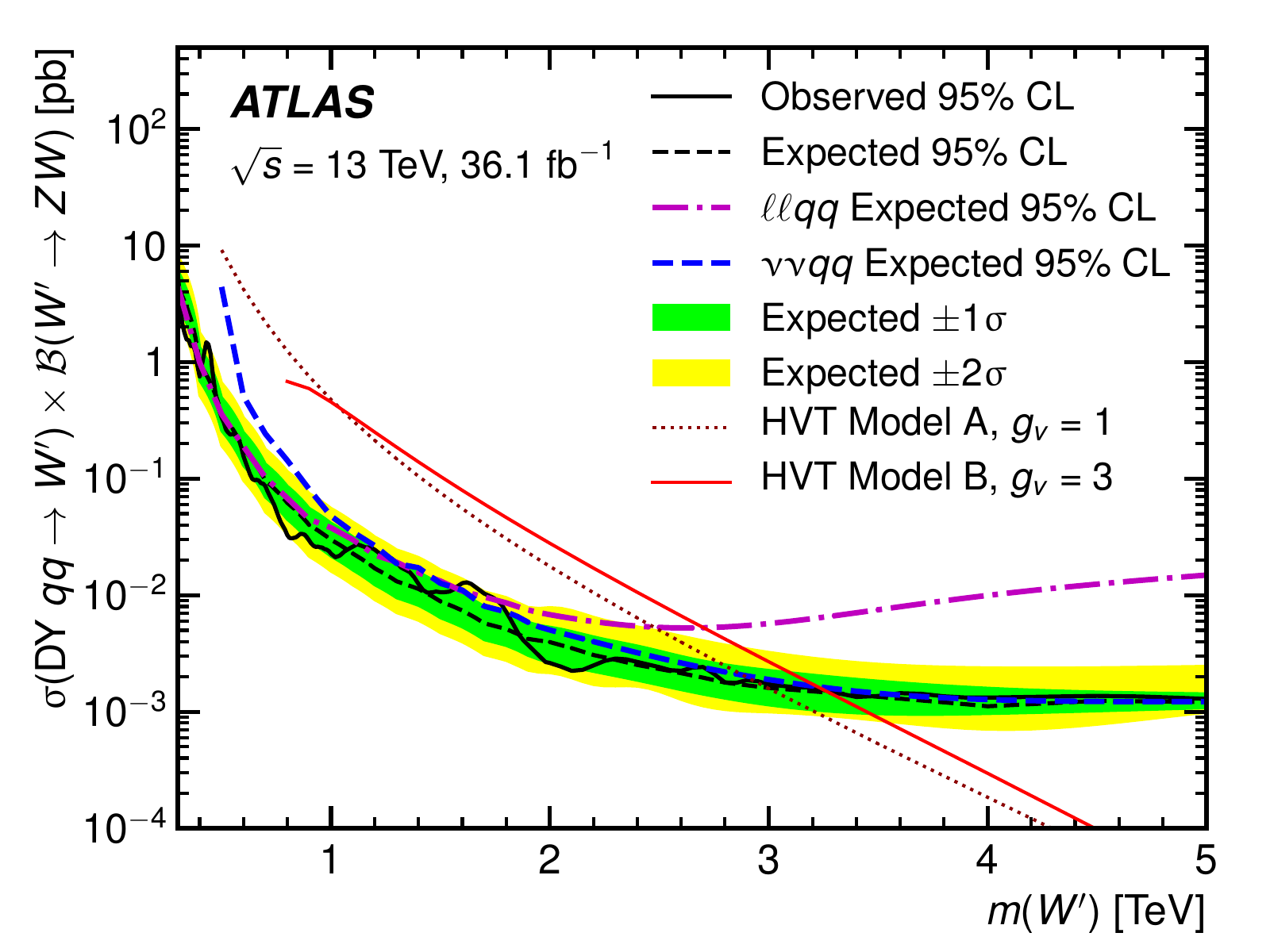}}
  \subfigure[VBF $W'\to ZW$]{\includegraphics[width=0.495\textwidth]{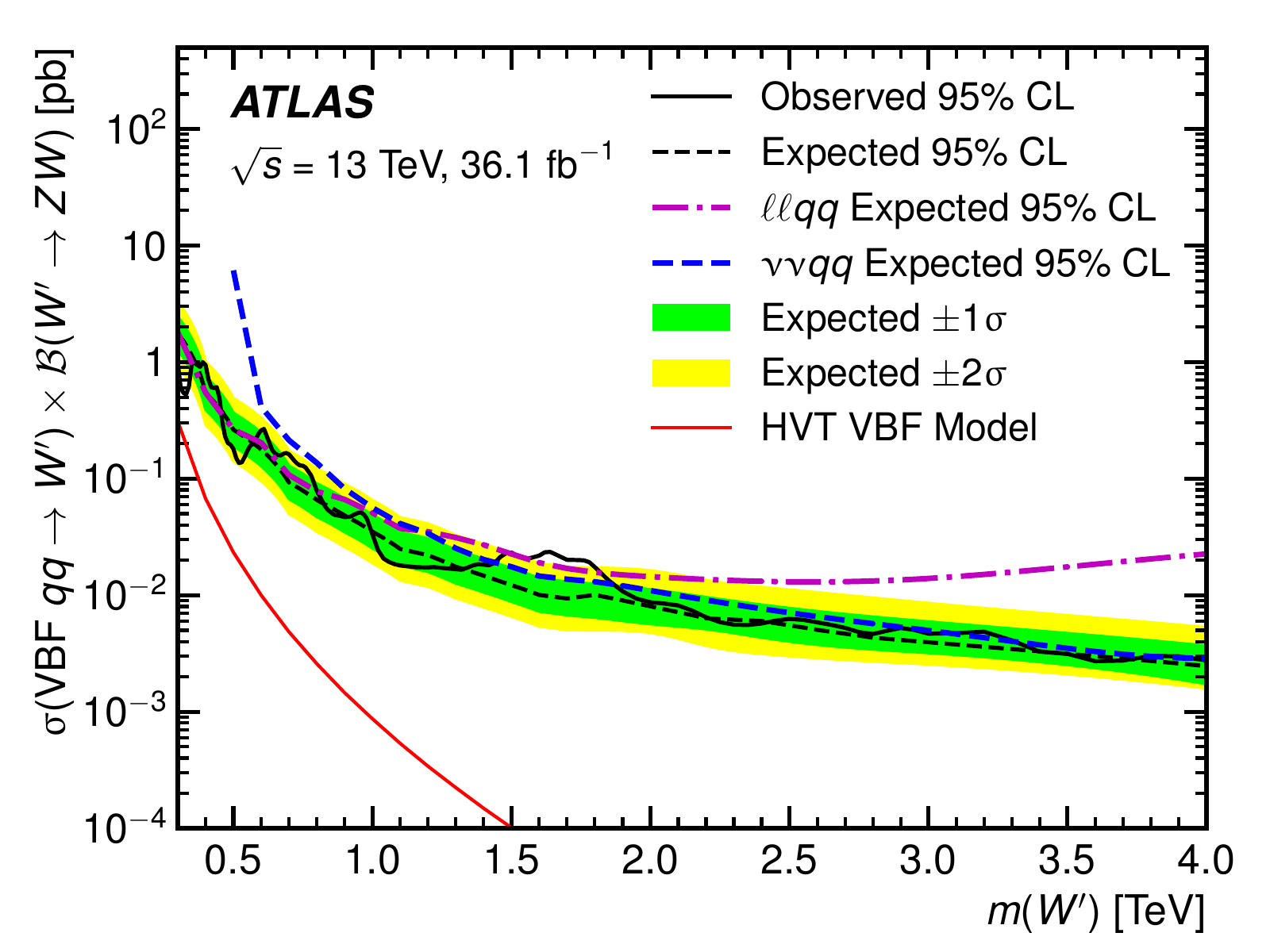}}  \\
  \caption{\label{fig:limits_HVT}
   Observed (black solid curve) and expected (black dashed curve) 95\% CL upper limits on $\sigma\times\mathcal{B}(W'\to ZW)$  at $\sqrt{s}=13$~TeV for the (a) DY and (b) VBF production of a $W'$ boson in the HVT model as a function of its mass, combining $\llqq$ and $\vvqq$ searches. Limits expected from individual searches (dashed curves in blue and magenta) are also shown for comparison.
Limits are calculated in the asymptotic approximation below 2~TeV and are obtained from pseudo-experiments above that.
Theoretical predictions are overlaid in (a) for HVT {\em Model A} and {\em Model B} and in (b) for HVT {\em VBF Model}.
The green (inner) and yellow (outer) bands represent the $\pm 1\sigma$ and $\pm 2\sigma$ uncertainty in the expected limits.}
\end{figure}

Figure~\ref{fig:limits_HVT} shows the limits on $\sigma\times\mathcal{B}(W'\to ZW)$ for DY and VBF production of a $W'$ boson in the HVT model. The observed limit ranges from 5.7~pb at 300~GeV to 1.3~fb at 5~TeV for DY production and from 0.98~pb at 300~GeV to 2.8~fb at 4 TeV for VBF production. The theoretical predictions of the HVT {\em Model A}, {\em Model B} and {\em VBF Model} are overlaid for comparison. The observed limits exclude an HVT $W'$ boson produced in the DY process lighter than 2.9~TeV for {\em Model A} and 3.2~TeV for {\em Model B}, while none of the HVT model space can be excluded for the VBF process with the current sensitivity of the analysis.
HVT models generally predict a near-by neutral vector boson ($Z'$) that decays into a $Z$ boson and a SM Higgs boson $h$. The $Z'$ production could contaminate both the signal and control regions of the HVT $W'$ search.  Assuming the $Z'\to Zh$ production having the same cross section as the observed upper limit on $\sigma\times\mathcal{B}(W'\to ZW)$, the impacts of the potential contamination to the $W'\to ZW$ search are found to be negligible.

\begin{figure}[h!bt]
\centering
  \subfigure[Bulk RS $\Graviton\to ZZ$, with $\kMPl=1$   ]{\includegraphics[width=0.495\textwidth]{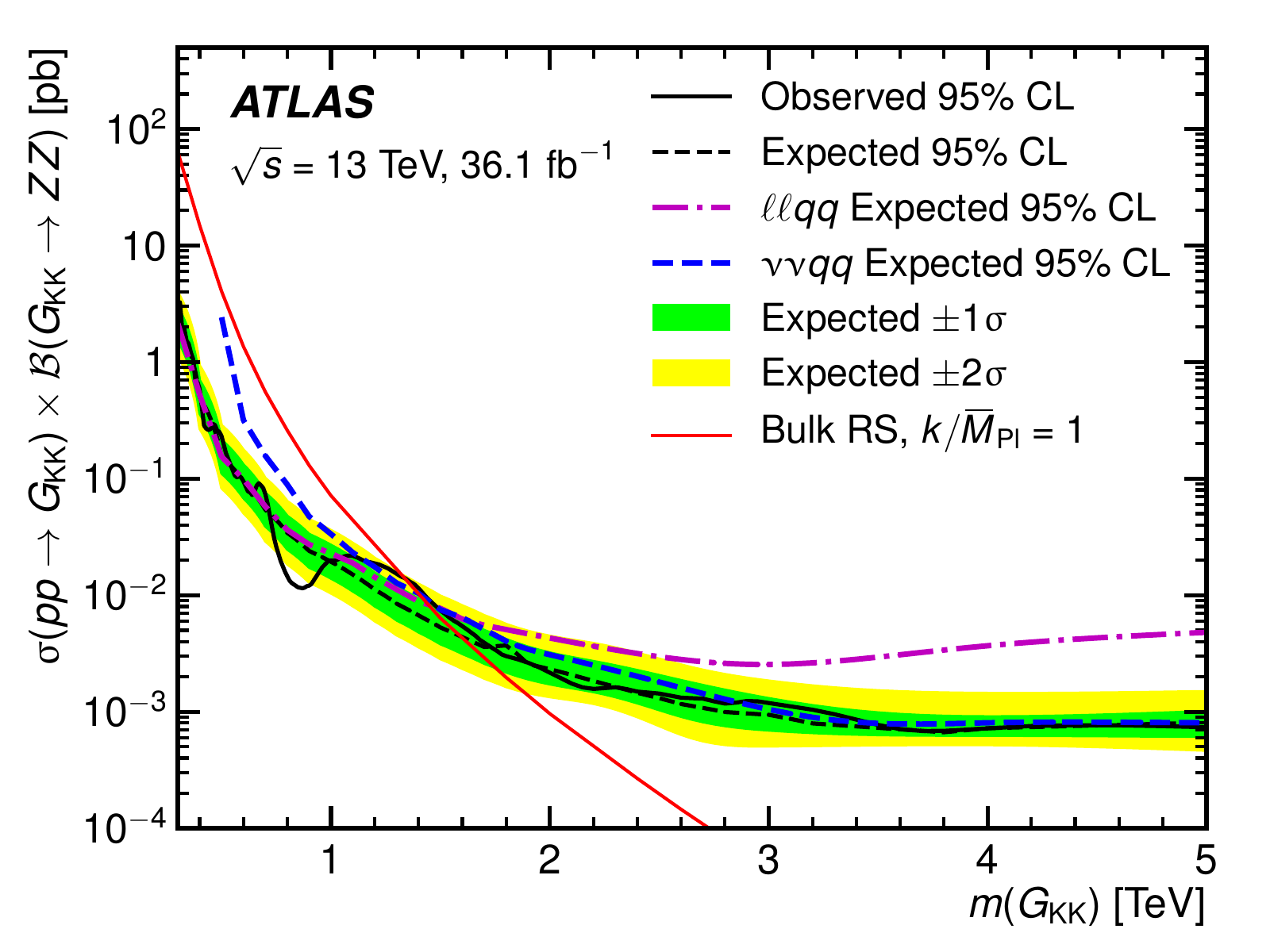}}
  \subfigure[Bulk RS $\Graviton\to ZZ$, with $\kMPl=0.5$ ]{\includegraphics[width=0.495\textwidth]{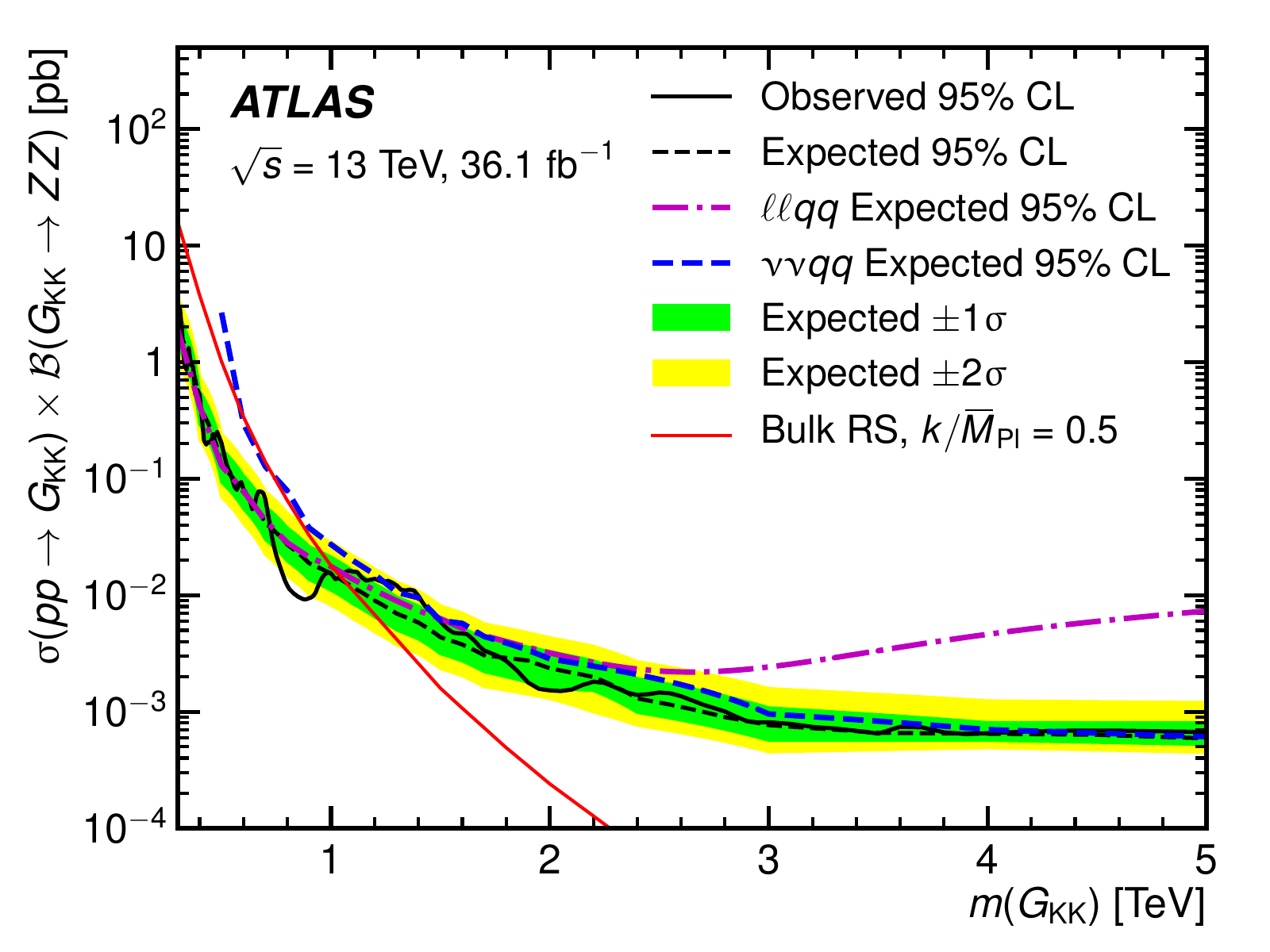}}
  \caption{\label{fig:limits_RSG}
   Observed (black solid curve) and expected (black dashed curve) 95\% CL upper limits on $\sigma\times\mathcal{B}(\Graviton\to ZZ)$ at $\sqrt{s}=13$~TeV for the production of a $\Graviton$ in the bulk RS model with couplings of (a) $\kMPl=1$ and (b) $\kMPl=0.5$ as a function of the graviton mass, combining $\llqq$ and $\vvqq$ searches. Limits expected from individual searches (dashed curves in blue and magenta) are also shown for comparison.
Limits are calculated in the asymptotic approximation below 2~TeV and are obtained from pseudo-experiments above that.
The theoretical predictions for $\sigma\times \mathcal{B}(\Graviton\to ZZ)$ as a function of resonance mass for a bulk RS graviton are also shown. The green (inner) and yellow (outer) bands represent the $\pm 1\sigma$ and $\pm 2\sigma$ uncertainty in the expected limits. }
\end{figure}

For $\Graviton\to ZZ$, limits are presented for two different couplings: $\kMPl=1$ and $\kMPl=0.5$, for masses between 300~GeV and 5~TeV, as shown in Figure~\ref{fig:limits_RSG}.
The observed limits on $\sigma\times\mathcal{B}(\Graviton\to ZZ)$ vary from 3.3~pb at 300~GeV to 0.74~fb at 5~TeV for the bulk RS model with $\kMPl=1$. The exclusion limit on the mass in this model is 1.3~(1.6) TeV for the observed~(expected) limit. Similar results are obtained for bulk RS model with $\kMPl=0.5$, with a mass exclusion upper limit of 1.0 TeV for both observed and expected limits.

These results are comparable or extend beyond previously published results~\cite{Aaboud:2016okv,Sirunyan:2017nrt}.

\subsection{Effects of systematic uncertainties}
The effects of systematic uncertainties are studied for hypothesised signals using the signal-strength parameter $\mu$.
 The relative uncertainties in the best-fit $\mu$ value from the leading
sources of systematic uncertainty
are shown in Table~\ref{tab:np_ranking} for ggF $H \to ZZ$ production with $m(H)=600$~GeV and 1.2~TeV. Apart from the statistical uncertainties in the data, the uncertainties with the largest impact on the sensitivity of the searches are from the size of the MC samples, measurements of small-$R$ and large-$R$ jets, \met\ measurement, background modelling and luminosity.
For signals with higher mass, the data statistical uncertainty becomes dominant.
The effects of systematic uncertainties for the other searches are similar to those shown for the ggF $H \to ZZ$ search.

\begin{table}[!ht]
\caption{\label{tab:np_ranking}
  The relative uncertainties from the leading uncertainty sources in the best-fit signal-strength parameter $\mu$ of hypothesised signal production of ggF $H\to ZZ$ with $m(H)=600$~GeV and $m(H)=1.2$~TeV. For this study, the $H\to ZZ$ production cross section is assumed to be 95~fb at 600~GeV and 13~fb at 1.2~TeV, corresponding to approximately the expected median upper limits at these two mass values.
}
\centering
\begin{tabular}{lcclc}
  \toprule
  \multicolumn{2}{c}{$m(H)=600$~GeV} &\hspace*{1.0cm}& \multicolumn{2}{c}{$m(H)=1.2$~TeV}\\  \midrule
  Uncertainty source    & $\Delta\mu/\mu$ [\%] && Uncertainty source    & $\Delta\mu/\mu$ [\%]     \\  \midrule

  Pseudo-data statistics           & 36   &&  Pseudo-data statistics            & 41    \\
  Total systematics                & 33   &&  Total systematics                 & 29    \\
  \midrule
  MC statistics                    & 20   &&  Large-$R$ jet                     & 20    \\
  Large-$R$ jet                    & 16   &&  Background modelling              & 13    \\
  \met\ uncertainties              & 13   &&  MC statistics                     & 13    \\
  Small-$R$ jet                    & 11   &&  Luminosity                        & 6.5   \\
  Background modelling             & 9.6  &&  Small-$R$ jet                     & 5.9   \\
  Luminosity                       & 9.1  &&  Leptons                           & 3.9   \\

  \bottomrule
\end{tabular}
\end{table}

%% file: tex/conclusion.tex
Searches for heavy resonances decaying into  $ZZ$ or $ZW$  are performed using proton--proton collision data produced at $\sqrt{s}=13$~TeV and recorded by the ATLAS detector at the LHC in 2015 and 2016. The data correspond to a combined integrated luminosity of \lumi. The searches explore the final states with one $Z$ boson decaying either into a pair of charged leptons or into a pair of neutrinos,  and the other $Z$ boson or the $W$ boson decaying into a pair of quarks, identified either as two separate jets or as one large-radius jet. Moreover, the searches are performed in two event categories, targeting VBF and non-VBF production of the resonances.

The data are found to be consistent with the SM background predictions and no evidence of heavy resonance production is observed.
Upper limits on the production cross section times branching ratio, $\sigma\times\mathcal{B}(X\to ZV)$, as a function of the resonance mass are derived at 95\% CL.
In the case of a heavy Higgs boson, upper limits on $\sigma\times\mathcal{B}(H\to ZZ)$ range from 1.7~(0.42) pb at $m(H)=300$~GeV to 1.4~(1.1)~fb at $m(H)=3$~TeV for the ggF~(VBF) production process.
For the phenomenological heavy-vector-triplet benchmark {\em Model A}~({\em Model B}) with coupling constant $g_{V}=1$~($g_{V}=3$),
a spin-1 vector triplet produced via the Drell--Yan process is excluded for $m(W^{\prime}) < 2.9~(3.2)$~TeV.
Upper limits on $\sigma\times\mathcal{B}(W'\to ZW)$ are also set for an HVT {\em VBF Model} with values of 0.98~pb at $m(W^{\prime})=300$~GeV and 2.8~fb at $m(W^{\prime})=4$~TeV.
For the bulk Randall--Sundrum model with $\kMPl=1$~($\kMPl=0.5$), a spin-2 Kaluza--Klein graviton is excluded for $m(\Graviton) < 1.3$~(1.0)~TeV.

These results extend previously published results~\cite{Aaboud:2016okv} obtained from the combinations of $\nu\nu qq,\, \ell\nu qq,\, \ell\ell qq$ and $qqqq$ final states based on a data sample of 3.2~fb$^{-1}$ at $\sqrt{s}=13$~TeV recorded in 2015.

%% file: acknowledgements/Acknowledgements.tex
% Acknowledgements for papers with collision data
% Version 16-Nov-2017

% Standard acknowledgements start here
%----------------------------------------------
We thank CERN for the very successful operation of the LHC, as well as the
support staff from our institutions without whom ATLAS could not be
operated efficiently.

We acknowledge the support of ANPCyT, Argentina; YerPhI, Armenia; ARC, Australia; BMWFW and FWF, Austria; ANAS, Azerbaijan; SSTC, Belarus; CNPq and FAPESP, Brazil; NSERC, NRC and CFI, Canada; CERN; CONICYT, Chile; CAS, MOST and NSFC, China; COLCIENCIAS, Colombia; MSMT CR, MPO CR and VSC CR, Czech Republic; DNRF and DNSRC, Denmark; IN2P3-CNRS, CEA-DRF/IRFU, France; SRNSF, Georgia; BMBF, HGF, and MPG, Germany; GSRT, Greece; RGC, Hong Kong SAR, China; ISF, I-CORE and Benoziyo Center, Israel; INFN, Italy; MEXT and JSPS, Japan; CNRST, Morocco; NWO, Netherlands; RCN, Norway; MNiSW and NCN, Poland; FCT, Portugal; MNE/IFA, Romania; MES of Russia and NRC KI, Russian Federation; JINR; MESTD, Serbia; MSSR, Slovakia; ARRS and MIZ\v{S}, Slovenia; DST/NRF, South Africa; MINECO, Spain; SRC and Wallenberg Foundation, Sweden; SERI, SNSF and Cantons of Bern and Geneva, Switzerland; MOST, Taiwan; TAEK, Turkey; STFC, United Kingdom; DOE and NSF, United States of America. In addition, individual groups and members have received support from BCKDF, the Canada Council, CANARIE, CRC, Compute Canada, FQRNT, and the Ontario Innovation Trust, Canada; EPLANET, ERC, ERDF, FP7, Horizon 2020 and Marie Sk{\l}odowska-Curie Actions, European Union; Investissements d'Avenir Labex and Idex, ANR, R{\'e}gion Auvergne and Fondation Partager le Savoir, France; DFG and AvH Foundation, Germany; Herakleitos, Thales and Aristeia programmes co-financed by EU-ESF and the Greek NSRF; BSF, GIF and Minerva, Israel; BRF, Norway; CERCA Programme Generalitat de Catalunya, Generalitat Valenciana, Spain; the Royal Society and Leverhulme Trust, United Kingdom.

The crucial computing support from all WLCG partners is acknowledged gratefully, in particular from CERN, the ATLAS Tier-1 facilities at TRIUMF (Canada), NDGF (Denmark, Norway, Sweden), CC-IN2P3 (France), KIT/GridKA (Germany), INFN-CNAF (Italy), NL-T1 (Netherlands), PIC (Spain), ASGC (Taiwan), RAL (UK) and BNL (USA), the Tier-2 facilities worldwide and large non-WLCG resource providers. Major contributors of computing resources are listed in Ref.~\cite{ATL-GEN-PUB-2016-002}.
%----------------------------------------------

%% file: atlas_authlist.tex
% ATLAS Collaboration author list
% Data extracted on 06-Jul-2017 for paper reference EXOT-2016-29
\begin{flushleft}
{\Large The ATLAS Collaboration}

\bigskip

M.~Aaboud$^\textrm{\scriptsize 137d}$,
G.~Aad$^\textrm{\scriptsize 88}$,
B.~Abbott$^\textrm{\scriptsize 115}$,
O.~Abdinov$^\textrm{\scriptsize 12}$$^{,*}$,
B.~Abeloos$^\textrm{\scriptsize 119}$,
S.H.~Abidi$^\textrm{\scriptsize 161}$,
O.S.~AbouZeid$^\textrm{\scriptsize 139}$,
N.L.~Abraham$^\textrm{\scriptsize 151}$,
H.~Abramowicz$^\textrm{\scriptsize 155}$,
H.~Abreu$^\textrm{\scriptsize 154}$
% ...
\bigskip
\\
% ...
$^{12}$ Institute of Physics, Azerbaijan Academy of Sciences, Baku, Azerbaijan\\
% ...
$^{88}$ CPPM, Aix-Marseille Universit{\'e} and CNRS/IN2P3, Marseille, France\\
% ...
$^{115}$ Homer L. Dodge Department of Physics and Astronomy, University of Oklahoma, Norman OK, United States of America\\
% ...
$^{119}$ LAL, Univ. Paris-Sud, CNRS/IN2P3, Universit{\'e} Paris-Saclay, Orsay, France\\
% ...
$^{137}$ $^{(a)}$ Facult{\'e} des Sciences Ain Chock, R{\'e}seau Universitaire de Physique des Hautes Energies - Universit{\'e} Hassan II, Casablanca; $^{(b)}$ Centre National de l'Energie des Sciences Techniques Nucleaires, Rabat; $^{(c)}$ Facult{\'e} des Sciences Semlalia, Universit{\'e} Cadi Ayyad, LPHEA-Marrakech; $^{(d)}$ Facult{\'e} des Sciences, Universit{\'e} Mohamed Premier and LPTPM, Oujda; $^{(e)}$ Facult{\'e} des sciences, Universit{\'e} Mohammed V, Rabat, Morocco\\
% ...
$^{139}$ Santa Cruz Institute for Particle Physics, University of California Santa Cruz, Santa Cruz CA, United States of America\\
% ...
$^{151}$ Department of Physics and Astronomy, University of Sussex, Brighton, United Kingdom\\
% ...
$^{154}$ Department of Physics, Technion: Israel Institute of Technology, Haifa, Israel\\
$^{155}$ Raymond and Beverly Sackler School of Physics and Astronomy, Tel Aviv University, Tel Aviv, Israel\\
% ...
$^{161}$ Department of Physics, University of Toronto, Toronto ON, Canada\\
% ...
$^{a}$ Also at Department of Physics, King's College London, London, United Kingdom\\
$^{b}$ Also at Institute of Physics, Azerbaijan Academy of Sciences, Baku, Azerbaijan\\
$^{c}$ Also at Novosibirsk State University, Novosibirsk, Russia\\
$^{d}$ Also at TRIUMF, Vancouver BC, Canada\\
$^{e}$ Also at Department of Physics {\&} Astronomy, University of Louisville, Louisville, KY, United States of America\\
% ...
$^{*}$ Deceased
\end{flushleft}

% Created with xml2latex.py